\documentclass[12pt,a4paper]{report}
\usepackage{amsmath,epsfig,amssymb,amsthm,subfigure}
\usepackage{float}
\usepackage{eso-pic}
\usepackage{graphicx,graphics,color}

\newcommand\WatermarkPicture{%
   \put(0,0){%
   \parbox[b][\paperheight]{\paperwidth}{%
     \vfill
     \centering
     \includegraphics[width=100pt,keepaspectratio]{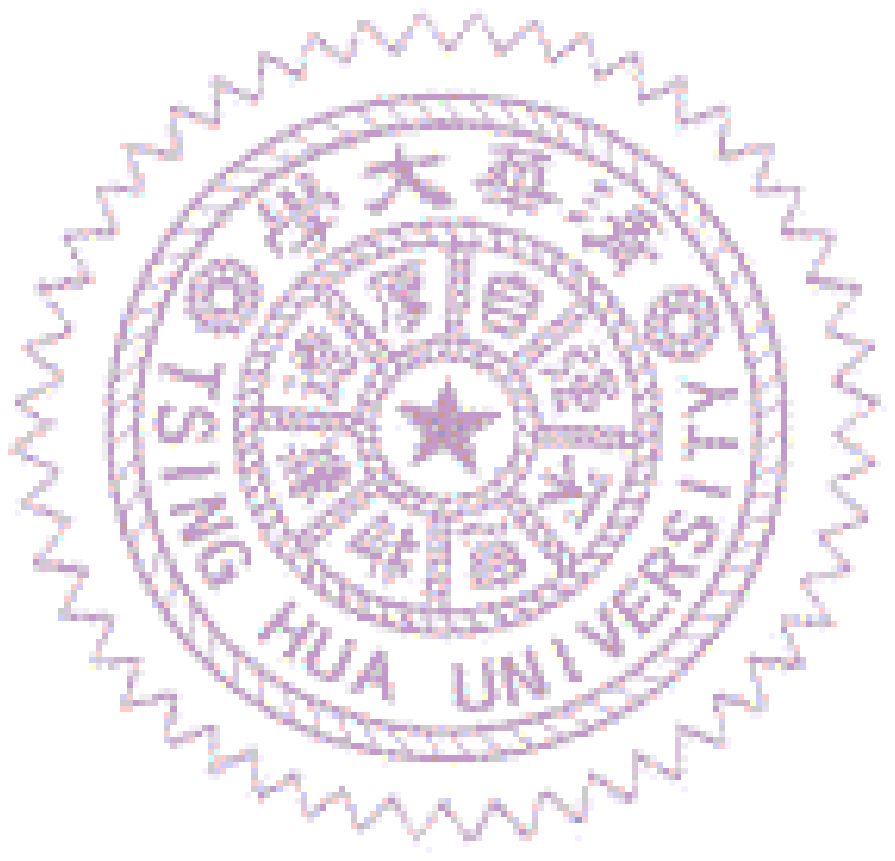}%
     \vfill
     }
   }
}

\title{Two-Way Training Design for Discriminatory Channel Estimation in Wireless MIMO Systems}
\author{Prepared by Chao-Wei Huang\\ \\
        Directed by Prof. Yao-Win Peter Hong\\ \\ \\ \\
        In Partial Fulfillment of the Requirements\\ \\
        for the Degree of\\
        Master of Science \\ \\ \\
        Institute of Communications Engineering\\ \\
        National Tsing Hua University \\ \\
        Hsinchu, Taiwan 30013, R.O.C.\\ \\
        E-mail: cwhuang@erdos.ee.nthu.edu.tw\\
        \date{July, 2011}
        }

\newtheorem{Prop}{Proposition}

\def\ie{{\it i.e.,}}
\def\etc{{\it etc}}
\def\iid{{\it i.i.d.}}
\def\eg{{\it e.g. }}
\def\Nt{{N_t}}
\def\NL{{N_L}}
\def\NU{{N_U}}
\def\cE{\mathcal{E}}
\def\cP{\mathcal{P}}
\def\bH{{\mathbf{H}}}     

\def\bHd{{\mathbf{H}_d}}
\def\bHu{{\mathbf{H}_u}}
\def\bG{{\mathbf{G}}}
\def\bHdt{{\mathbf{H}_{d,t}}}
\def\hbHd{{\mathbf{\widehat{H}}_d}}
\def\hbHdt{{\mathbf{\widehat{H}}_{d,t}}}
\def\hbHu{{\mathbf{\widehat{H}}_u}}

\def\sHd2{\sigma_{H_d}^2}
\def\sHu2{\sigma_{H_u}^2}
\def\sG2{\sigma_G^2}
\def\bbC{\mathbb{C}}
\def\bI{\mathbf{I}}

\def\bA{{\mathbf{A}}}
\def\sa2{{\sigma_a^2}}
\def\bNHd{{\mathbf{N}_{\widehat{H}_d}}}
\def\bNHdt{{\mathbf{N}_{\widehat{H}_{d,t}}}}
\def\bNHdH{{\mathbf{N}^H_{\widehat{H}_d}}}
\def\bNHdHt{{\mathbf{N}^H_{\widehat{H}_{d,t}}}}

\def\bV{{\mathbf{V}}}
\def\sv2{{\sigma_v^2}}

\def\bX{{\mathbf{X}}}
\def\bC{{\mathbf{C}}}
\def\bY{\mathbf{Y}}
\def\bW{\mathbf{W}}
\def\sw2{\sigma_w^2}
\def\bWt{{\mathbf{\widetilde{W}}}}
\def\swt2{\sigma_{\tilde{w}}^2}
\def\bPave{P_{ave}}
\def\bPL{\bar{P}_L}
\def\bPt{\bar{P}_t}
\def\tgam{{\tilde{\gamma}}}   

\psfull
\begin{document}
\AddToShipoutPicture{\WatermarkPicture} 
\baselineskip 12pt \baselineskip 24pt \maketitle
\pagenumbering{roman}
\addcontentsline{toc}{chapter}{Abstract}

\chapter*{Abstract}

This work examines the use of two-way training in multiple-input
multiple-output (MIMO) wireless systems to discriminate the channel
estimation { performances between a legitimate receiver (LR) and an unauthorized
receiver (UR). This thesis extends upon the previously proposed discriminatory channel
estimation (DCE) scheme that allows only the transmitter to send training signals.
The goal of DCE is to minimize the channel estimation error at LR while requiring the channel estimation error at UR to remain beyond a certain level. If the training signal is sent only by the transmitter, the performance discrimination between LR and UR will be limited since the training signals help both receivers perform estimates of their downlink channels. In this work, we consider instead the two-way training methodology that allows both the transmitter and LR to send training signals. In this case, the training signal sent by LR helps the transmitter obtain knowledge of the transmitter-to-LR channel, but does not help UR estimate its downlink channel (i.e., the transmitter-to-UR channel).
With transmitter knowledge of the estimated transmitter-to-LR channel, artificial noise (AN) can then be embedded in the null space of the transmitter-to-LR channel to disrupt UR's channel estimation without severely degrading the channel estimation at LR.
Based on these ideas, two-way DCE training schemes are developed for both reciprocal and non-reciprocal channels. The optimal power allocation
between training and AN signals is devised under both average and individual power constraints. Numerical results are provided to
demonstrate the efficacy of the proposed two-way DCE training schemes.}

\newpage
\addcontentsline{toc}{chapter}{Contents} \tableofcontents
\listoffigures

\newpage
\pagenumbering{arabic} \setcounter{page}{1}

\chapter{Introduction}

{Secrecy in wireless communications has {been an important problem over the years} due to the broadcast nature of the
wireless medium. In the past, these issues have mostly been
addressed using cryptography in the application layer. However,
recent studies on information-theoretic secrecy provide an
alternative to achieve these tasks through coding and modulation in
the physical layer. In the context of physical layer secrecy, one is often interested in
deriving the so-called secrecy capacity, which is the rate achievable
with vanishing error probability at the legitimate receiver (LR) and
vanishing equivocation rate at the unauthorized receiver (UR). In particular, secrecy capacity has
been derived for single-input single-output (SISO) systems in
\cite{secrecy.capacity} and for multiple-input multiple-output
(MIMO) systems in \cite{Khisti}. The results show that secrecy
capacity {can generally} be increased by enlarging the difference between the
effective channel qualities of LR and UR.
While most works on physical layer secrecy focus on optimal coding schemes to achieve secrecy capacity in the data transmission phase, our goal is to exploit signal processing methods to enlarge the differences between the quality of the two channels. In particular, this is done from a channel estimation aspect, following the so-called {\it discriminatory channel estimation} (DCE) methodology proposed previously in \cite{ChangChiangHongChi_TSP2010}.


Specifically, DCE is a training strategy that utilizes artificial noise (AN) to disrupt UR's reception while sending training signals to LR.
In this case, AN must be placed in the null space of the transmitter-to-LR channel to minimize its effect on LR. However, this requires transmitter knowledge of the channel, which is typically obtained through feedback from LR. In the original DCE scheme, only the transmitter is allowed to send training signals. In this case, increasing the training power helps improve the channel estimate at LR and allows for more effective use of AN at the transmitter. However, this also helps UR obtain a better channel estimate and, thus, the amount of power used for training must be limited. To improve the channel estimate at LR while confining the performance of UR to a certain level, multiple stages of feedback and retraining must be employed. In this case, training and AN power is increased as the transmitter knowledge of the channel improves through multiple stages.
Yet, the training overhead and complexity required to optimize training over multiple stages limits its application in practice.

The main contribution of this thesis is to propose new and efficient DCE schemes using the two-way training methodology. Here, training signals will also be sent by LR and transmitter knowledge of the channel will be obtained by performing channel estimation at the transmitter. Notice that the original DCE scheme assumes that channel feedback with infinite resolution is provided from LR, which is not achievable in practice.}
When the channel is reciprocal, $\eg$, in time-division multiplexing
(TDD) systems, the channel state information (CSI) can be obtained at
the transmitter by {sending} pilot signals from the receiver.
{Two-way training schemes have been studied for conventional point-to-point links} in \cite{twoway.SIMO,twoway.MISO,twoway.ECHOMIMO}
to obtain the CSI at both the receiver and the transmitter without the
use of feedback. In this work, we adopt the concept of two-way
training into the design to increase the efficiency of the DCE
scheme. In reciprocal channels, the proposed two-way DCE scheme uses
reverse training to provide CSI at the transmitter and
forward training with AN to achieve different channel estimation
performances at LR and UR. When the channel
is non-reciprocal, $\eg$, in frequency-division multiplexing (FDD)
systems, {the downlink and uplink channels between the transmitter
and LR would not be identical. In this case, an additional training phase is needed,
where the transmitter first broadcasts
a randomly generated signal to LR, which then echoes the signal back to the transmitter.
The echoed signal contains information of both the downlink and uplink channels and can be combined with the reverse training signal to estimate the desired transmitter-to-LR channel ($\ie$ the downlink channel).} Compared to the multi-stage feedback-and-retraining DCE scheme in
\cite{ChangChiangHongChi_TSP2010}, {the proposed two-way training scheme
drastically decreases the overall training overhead and design complexity.

To optimize the performance of the proposed two-way training scheme, we derive the optimal power allocation between the
training  and AN by} solving an optimization problem that
aims to minimize the channel estimation error at the LR subject to a
lower limit constraint on the channel estimation error at the UR.
In the reciprocal case, the analytical result shows that the problem
of finding the optimal training and AN powers
reduces to a one-variable problem which can be solved by a simple
line search. However, in the non-reciprocal case, the power allocation
problem {is not easily solved since the estimation error expression is much more complex. Therefore, we}
instead resort to an approximate solution
by using the monomial approximation and condensation method
\cite{Tutorial_GP} in the field of geometric programming (GP).
Numerical results show that the proposed DCE design can
effectively discriminate {the channel estimation} and the
data detection performances at the LR and UR.

The remainder of the thesis is organized as follows. In Chapter \ref{ch2.system}, we first introduce the
wireless MIMO system model considered in this work and provide a general description of the DCE scheme.
For the case with reciprocal channels, the training strategy is described in Chapter \ref{ch3.Recip} and the optimal power allocation is derived in Chapter \ref{ch4.Recip.Anal}. Similarly, for the case with nonreciprocal channels, the training strategy is described in Chapter \ref{ch5.NonRecip} and the optimal power allocation is derived in \ref{ch6.Nonrecip.Anal}. Numerical results are provided in Chapter \ref{ch7.simulation} and, finally, a conclusion is given in Chapter \ref{ch8.conclusion}.

\chapter{System Model} \label{ch2.system}

\begin{figure}[t]
\centering
{\includegraphics[scale=0.8]{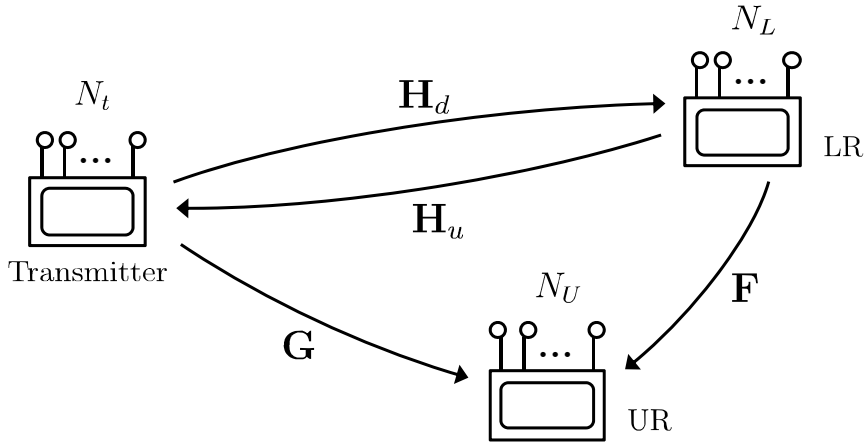}}
\caption{Diagram of a wireless MIMO system consisting of a transmitter, a legitimate receiver
(LR) and an unauthorized receiver (UR).}
\label{fig.ch2.sysmodel}
\end{figure}

Consider a wireless MIMO system {that consists} of a transmitter,
a legitimate receiver (LR), and an unauthorized receiver (UR), as shown in
Fig. \ref{fig.ch2.sysmodel}. We assume that the transmitter, LR, and UR are
equipped with $N_t$, $N_L$ and $N_U$ antennas, respectively. The channels of
LR and UR remain constant during one transmission block, which consists of a
training phase and a data transmission phase.
The goal is to prevent the UR to extract information from its received signal. Instead of focusing the data transmission, we
propose to achieve this task from a channel estimation perspective and devise two-way training schemes following the DCE methodology that
enables LR to perform an accurate estimate of the channel while
disrupting the channel estimation performance at UR.

Let the downlink channel from the
transmitter to LR be denoted by $\bHd \in \bbC^{\Nt \times \NL}$ and the uplink
channel from LR to the transmitter be denoted by $\bHu \in \bbC^{\NL \times \Nt}$.
In the following chapters, we consider separately two different
channel models, i.e., the reciprocal channel model and the non-reciprocal channel model.
In both cases, the proposed two-way
training scheme for DCE can be generally divided into two steps as
described below.

{\bf Step I}: The aim of Step I is to allow the transmitter to obtain an estimate of the downlink channel.
Different from  \cite{ChangChiangHongChi_TSP2010}, where a noiseless feedback channel is required, we allow the transmitter to estimate the downlink channel itself through the exchange of training signals between the transmitter and UR. This channel knowledge will be used in designing the forward training signal
in order to discriminate the channel estimation performances between LR and UR.
Different training strategies are required under different channel models to achieve the
downlink channel estimation at the transmitter. Detailed descriptions for
the reciprocal case and non-reciprocal case are given in Chapters
\ref{ch3.Recip} and \ref{ch5.NonRecip}, respectively.

{\bf Step II}: After obtaining the downlink channel estimate in Step I, the transmitter next
sends a training signal along with AN to degrade the channel estimation
performance at UR. Specifically, by assuming that $\Nt>\NL$, the
forward training signal is given by
\begin{equation}\label{eq.forward.training.AN}
\bX_{t}=\sqrt{\frac{\cP_F\tau_F}{\Nt}}\bC_{t}+\bA\bNHdH,
\end{equation}
where $\bC_{t}\in \bbC^{\tau_F\times \Nt}$ is the pilot matrix with
$\text{Tr}(\bC_{t}^H\bC_{t})=\Nt$, $\cP_F$ is the forward training power, and $\tau_F$ training length. For ease of notation, we define $\cE_F\triangleq \cP_F \tau_F$ as the forward training energy.
$\bA \in \bbC^{\tau_F \times(\Nt-\NL)}$ is the AN matrix of
which each entry is $\iid$ Gaussian with zero mean and variance $\sa2$ and
$\bNHd\in\bbC^{\Nt\times(\Nt-\NL)}$ is a matrix whose column
vectors form an orthonormal basis for the left null space of $\hbHd$, that is, $\bNHdH\hbHd=\mathbf{0}_{(\Nt-\NL)\times \NL}$ (i.e., the
($\Nt-\NL$) by $\NL$ zero matrix) and $\bNHdH\bNHd=\bI_{\Nt-\NL}$.
Notice from (\ref{eq.forward.training.AN}) that AN is superimposed on the training signal and placed
in the left null space of $\hbHd$ to minimize its interference on LR.
The received signals of the LR and UR are respectively given by
\begin{align}\label{eq.receive.signal.LR}
\bY_{L}&=\sqrt{\frac{\cE_F}{\Nt}}\bC_{t}\bHd+\bA\bNHdH\bHd+\bW\\
\bY_{U}&=\sqrt{\frac{\cE_F}{\Nt}}\bC_{t}\bG+\bA\bNHdH\bG+\bV \label{eq.receive.signal.UR}
\end{align}
where $\bG\in\bbC^{\Nt\times\NU}$ is the channel matrix from the transmitter to UR,
and $\bW\in\bbC^{\tau_F\times\NL}$ and $\bV\in\bbC^{\tau_F\times \NU}$ are the additive
white noise matrices at LR and UR, respectively. Each entry of $\bG$ is assumed to be
$\iid$ distributed with zero mean and variance equal to $\sG2$. Elements of both
$\bW$ and $\bV$ are assumed to be  $\iid$ random variables with zero mean and variances
respectively equal to $\sw2$ and $\sv2$.

In the following chapters, we describe the training strategies and examine the corresponding channel estimation performances at the transmitter and the receivers during each stage of the process. The optimal power allocation between training and AN signals are derived to achieve discrimination between the channel estimation performances at LR and  UR.

\chapter{Two-Way Training Strategy for Reciprocal Channels}\label{ch3.Recip}

{In this chapter, we consider the case where the channel between the transmitter and the LR
is reciprocal, which means that the downlink and uplink channels are symmetric.
In this case, we define the downlink channel matrix as $\bHd\triangleq \bH$ and the uplink
channel matrix as $\bHu \triangleq \bH^T$. With channel reciprocity, the transmitter
can obtain an estimate of the downlink channel by taking the transpose of the channel matrix obtained through reverse training, i.e., training based on signals sent from LR to the transmitter. In this case, DCE is effectively achieved using only two stages, i.e., the reverse and the forward training phases. The operations are detailed below.

{\bf Reverse Training}: In the reverse training stage, LR} first sends a training signal,
denoted by $\bX_L\in\bbC^{\tau_R \times \NL}$, to enable channel
estimation at the transmitter. Specifically,
the reverse training signal $\bX_L$ is given by
\begin{equation} \label{equ.training.LR}
\bX_L=\sqrt{\frac{\cP_R\tau_R}{N_L}}\bC_L,
\end{equation}
where the pilot matrix $\bC_L$ satisfies $\bC^H_L\bC_L=\bI_\NL$
(the $N_L$ by $N_L$ identity matrix), and $\cP_R$ and $\tau_R$ represent
the transmission power and training interval, respectively. For ease of use later,
we define the reverse training as $\cE_R\triangleq \cP_R \tau_R$.
The received signal at the transmitter is given by
\begin{equation}\label{equ.rec.tx}
\bY_t=\bX_L\bH^T+\bWt,
\end{equation}
where each element of $\bH$ is assumed to be independent and
identically distributed (i.i.d.) random variable with zero mean and
variance equal to $\sigma_H^2$, and $\bWt \in \bbC^{\tau_R \times
\Nt}$ is the additive white noise matrix with each element having
zero mean and variance $\swt2$. By the help of reverse training,
the channel estimate of $\bH$, denoted by $\mathbf{\widehat{H}}$, can
be obtained at the transmitter

{\bf Forward Training}:
In the forward training stage, the transmitter superimposes AN on top of the training signal to
degrade the channel estimation performance at UR. With knowledge of the estimated downlink channel, i.e., $\mathbf{\widehat{H}}$, AN can be placed
in the left null space of $\mathbf{\widehat{H}}$ to minimize its interference
on the LR. The forward training signal is given in (\ref{eq.forward.training.AN}) where the pilot matrix $\bC_t$ satisfies
$\bC^H_t\bC_t=\bI_\Nt$.
And the received signals of LR and UR are respectively given in
(\ref{eq.receive.signal.LR}) and (\ref{eq.receive.signal.UR}). Note that
the notation $\hbHd$ in (\ref{eq.forward.training.AN}) and (\ref{eq.receive.signal.LR})
is replaced by $\mathbf{\widehat{H}}$.

In the next chapter, we analyze the channel estimation performances
of transmitter, LR and UR by assuming that all of them employ
linear minimum mean square error (LMMSE) criterion for
channel estimation \cite{Est_Theory}.
We then propose to judiciously allocate the training powers and the AN power
in reverse and forward training, aiming at discriminating between the channel
estimation performances of the LR and the UR.

\chapter{Optimal Power Allocation for DCE in Reciprocal Channels}\label{ch4.Recip.Anal}

\section{Channel Estimation Performance at Transmitter}
Due to channel reciprocity, the reverse training signals
sent by LR allow the transmitter to obtain an estimate of the downlink channel by taking the transpose of its estimate of the uplink channel. By employing the LMMSE estimator, the estimate of the channel matrix $\bH$ can be written as
\begin{align}
\widehat{\mathbf{H}}&=(\sigma_H^2\mathbf{X}_{L}^H(\sigma_H^2\mathbf{X}_{L}\mathbf{X}_{L}^H+
\sigma_{\tilde{w}}^2\mathbf{I}_{\tau_R})^{-1}\mathbf{Y}_{t})^T \notag \\&\triangleq
\bH+\Delta\bH \label{equ.est.rev.channel}
\end{align}
where $\Delta\bH\in\bbC^{\Nt\times \NL}$ stands for the estimation
error matrix. The covariance matrix of $\Delta\bH$ can be shown to
be \cite{Est_Theory}
\begin{align}
\mbox{E}\{\Delta\bH(\Delta\bH)^H\}\label{MSE.forward.channel.tx}
&=\NL\left(\frac{1}{\sigma_H^2}+\frac{\cE_R}{N_L\sigma_{\tilde{w}}^2}\right)^{-1}\bI_{\Nt}.
\end{align}

\section{Channel Estimation Performance at LR and UR}
To analyze the channel estimation performance of LR, let us write
(\ref{eq.receive.signal.LR}) as
\begin{align} \label{equ.rec.LR}
\bY_L&=\sqrt{\frac{\cE_F}{\Nt}}\bC_t\bH-\bA\mathbf{N}_{\hat{H}}^H\Delta\bH+\bW
\triangleq\bar{\mathbf{C}}\bH+\bar{\mathbf{W}},
\end{align}
where $\bar{\mathbf{C}}\triangleq \sqrt{\frac{\cE_F}{\Nt}}\bC_t$,
$\bar{\mathbf{W}}\triangleq -\bA\mathbf{N}_{\hat{H}}^H\Delta{\bH}+\bW$ and
the first equality is due to
$\mathbf{N}_{\hat{H}}^H\mathbf{\widehat{H}}=\mathbf{0}$. Denote
the channel estimate at LR by $\mathbf{\widehat{H}}_L$. The normalized mean
squared error (NMSE) of $\mathbf{\widehat{H}}_L$ under LMMSE criterion can be
shown to be \cite{Est_Theory}
\begin{align}\label{NMSE.forward.channel.LR}
\mbox{NMSE}_L&\triangleq
\frac{\mbox{Tr}\left(\mbox{E}\{(\bH-\mathbf{\widehat{H}}_L)(\bH-\mathbf{\widehat{H}}_L)^H\}\right)}
{\Nt\NL}\notag\\
&=\frac{\mbox{Tr}\left(\left(\mathbf{R}_H^{-1}
+\bar{\mathbf{C}}^H\mathbf{R}_{\bar{W}}^{-1}\bar{\mathbf{C}}\right)^{-1}\right)}{\Nt\NL},
\end{align}
where $\mbox{Tr}(\cdot)$ denotes the trace of a matrix,
$\mathbf{R}_H=\NL\sigma_H^2\bI_{\Nt}$ and
$\mathbf{R}_{\bar{W}}=\mbox{E}\{\mathbf{W}\bar{\mathbf{W}}^H \} $ is the
covariance matrix of $\bar{\mathbf{W}}$. According to the
independence between $\bA$ and $\bW$, the fact of
$\mathbf{N}_{\widehat{H}}^H\mathbf{N}_{\widehat{H}}=\bI_{\Nt-\NL}$ and
(\ref{MSE.forward.channel.tx}), it can be shown that
\begin{align}
\mathbf{R}_{\bar{W}}&=\left(\mbox{E}\{\lVert\mathbf{N}_{\widehat{H}}^H\Delta\bH\rVert^2\}
\sigma_a^2+\NL\sigma_w^2 \right)\bI_{\tau_F}\notag\\
&=\NL\left[(\Nt-\NL)\cdot \left(\frac{1}{\sigma_H^2}+
\frac{\cE_R}{\NL\sigma_{\tilde{w}}^2} \right)^{-1}\sigma_a^2 +
\sigma_w^2\right]\bI_{\tau_F}.\label{var.noise.LR}
\end{align}
Substituting (\ref{var.noise.LR}) into
(\ref{NMSE.forward.channel.LR}) yields
\begin{align}
\mbox{NMSE}_L
&=\frac{\mbox{tr}\left(\frac{1}{\NL\sigma_H^2}\bI_{\Nt}+\frac{\cE_F}{\Nt\NL}\frac{\bC^H_{t}\bC_t}
{(\Nt-\NL)\left(\frac{1}{\sigma_H^2}
+\frac{\cE_R}{\NL\sigma_{\tilde{w}}^2}\right)^{-1}\sigma_a^2+\sigma_w^2}\right)^{-1}}{\Nt\NL}\notag \\
&=\left(\frac{1}{\sigma_H^2}+\frac{{\cE_F}/{\Nt}}{(\Nt-\NL)\left(\frac{1}{\sigma_H^2}+
\frac{\cE_R}{\NL\sigma_{\tilde{w}}^2}\right)^{-1}\sigma_a^2+\sigma_w^2}
\right)^{-1}. \label{NMSE2.forward.channel.LR}
\end{align}
The NMSE performance of the UR can be analyzed in a similar way.
Specifically, one can show that the NMSE of estimating $\bG$ at the UR
is given by
\begin{align}\label{NMSE2.forward.channel.UR}
\mbox{NMSE}_U
&=\left(\frac{1}{\sigma_G^2}+\frac{{\cE_F}/{\Nt}}{(\Nt-\NL)\sigma_a^2\sigma_G^2+\sigma_v^2}\right)^{-1}.
\end{align}

\section{Optimal Power Allocation between Training and AN Signals}

Observing from \eqref{NMSE2.forward.channel.LR} and
\eqref{NMSE2.forward.channel.UR}, the added AN in forward training
can affect both the LR and UR's channel estimation performances. To
optimize LR's channel estimation performance while preventing the UR
from obtaining an accurate estimate of $\bG$, we propose to jointly
optimize the reverse training energy $\cE_R$, the forward training
energy $\cE_F$ and AN power $\sigma_a^2$ by considering the following
power allocation problem
\begin{align}\label{opt.prob.I1}
\min_{\cE_R,\cE_F,\sigma_a^2\geq0}&\mbox{NMSE}_L \\
\mbox{s.t.}~~~ &\mbox{NMSE}_U \geq \gamma, \notag\\
&\cE_R +\cE_R+(\Nt-\NL)\sigma_a^2\tau_F \leq \bPave(\tau_R+\tau_F),\notag\\
&\cE_R \leq \bPL \tau_R,\notag\\
&\cE_F+(\Nt-\NL)\sigma_a^2\tau_F \leq \bPt \tau_F,\notag
\end{align}
where we aim to minimize the LR's NMSE subject to a preset lower
limit $\gamma$ on the UR's NMSE, under an average power constraint
$\bPave$. Note that the LR and the transmitter also have their own
peak power constraints, i.e., $\bPL$ and $\bPt$.

{\bf Remark}: In the DCE scheme, it is desirable to keep
the forward training length as small as possible, $\ie$ equal to the number
of transmit antennas. Observing from (\ref{NMSE2.forward.channel.LR}),
(\ref{NMSE2.forward.channel.UR}) and the problem (\ref{opt.prob.I1}) and
assuming the average energy constraint and the individual energy constraint
on the transmitter and the LR are all fixed, as the forward training
length increases, it needs more AN energy to meet the same
lower limit value and thus less energy can be allocated to
the training signal. Different from the receiver's noise of which
the energy can be freely accumulated over time, it takes the system's
resource to maintain the AN's power.

To make all
constraints effective, we shall focus on the interesting case where
\begin{equation}\label{interesting case}
\max\{\bPL \tau_R,\ \bPt \tau_F\} \leq
\bPave(\tau_R+\tau_F) \leq \bPL \tau_R+\bPt \tau_F.
\end{equation}
Note that for the case where $\bPave(\tau_R+\tau_F) >\bPL \tau_R+\bPt \tau_F$,
the average power constraint becomes redundant and hence, the transmitter
and the LR simply transmit with its maximum power.
When $\bPave(\tau_R+\tau_F) < \bPL \tau_R$ and/or
$\bPave(\tau_R+\tau_F) < \bPt \tau_F$, one or both individual power
constraints become redundant. The solution for this case can be easily
obtained by following the derivations for the case of (\ref{interesting case})
\footnote{The proposition to be given for the case of (\ref{interesting case})
also describes the solution for the case of $\bPave(\tau_R+\tau_F) < \bPL \tau_R$
and/or $\bPave(\tau_R+\tau_F) < \bPt \tau_F$, by changing the condition
in (\ref{interesting condition}) to $0\leq \tgam \leq \min\{\bPt \tau_F,
\ \bPave(\tau_R+\tau_F)\}$ and setting the redundant individual power
constraint(s) to infinity.}.

On the other hand, it should be noted that the preset value
$\gamma$ should satisfy \cite{ChangChiangHongChi_TSP2010}
\begin{equation}\label{cond.gamma}
\left(\frac{1}{\sigma_G^2}+\frac{\min\{\bPt\tau_F,\bPave(\tau_R+\tau_F)\}}
{\Nt\sigma_v^2}\right)^{-1} \leq \gamma \leq \sigma_G^2,
\end{equation} since the left-hand-side term is the minimum
achievable NMSE of UR (when the transmitter does not use AN, i.e.,
$\sigma_a^2=0$), and the right-hand-side term stands for the worst
NMSE performance of UR, respectively. For ease of latter use, let us
define \begin{equation} \tgam \triangleq \left(
\frac{1}{\gamma}-\frac{1}{\sigma_G^2}\right)\Nt\sigma_v^2\geq 0.
\end{equation}
Then the condition in (\ref{cond.gamma}) reduces to
\begin{equation}\label{interesting condition}
0\leq \tgam \leq \bPt \tau_F.
\end{equation}

The power allocation problem in \eqref{opt.prob.I1} is a nonconvex
optimization problem involving three variables ($\cE_R,\cE_F,\sigma_a^2$).
However, it actually can be solved very efficiently.
We show in Appendix 9.1 the following proposition for problem
\eqref{opt.prob.I1}:
\begin{Prop}\label{prop.optpower.rec} Consider the power allocation problem in
\eqref{opt.prob.I1} with both \eqref{interesting case} and
\eqref{interesting condition} satisfied. If
\begin{equation} \label{condition u}
\mu \triangleq \NL
\left(\frac{\sigma_v^2\sigma_{\tilde{w}}^2}{\sigma_G^2\sigma_w^2}
-\frac{\sigma_{\tilde{w}}^2}{\sigma_H^2}\right) \notag
>\min\{\bPL \tau_R,\ \bPave(\tau_R+\tau_F)-\tgam\}
\end{equation}
then the optimal $(\cE_R,\cE_F,\sa2)$ of \eqref{opt.prob.I1} is
given by $\cE_R*=0$, $\cE_F^*=\tgam$ and $(\sa2)^*=0$ (i.e., no need of reverse
training and no need of AN in forward training). On the other hand,
if $\mu \leq \min\{\bPL \tau_R,\ \bPave(\tau_R+\tau_F)-\tgam\}$, the optimal
$\cE_R$ of \eqref{opt.prob.I1} can be obtained by solving the following
one-dimensional problem
\begin{align}\label{opt.prob.final1}
\cE_R^\star=\arg~\max_{\cE_R\geq 0}\ \ \ &\frac{(\NL\swt2+\sigma_H^2\cE_R)b(\cE_R)}{\NL\swt2+\sigma_H^2\cdot \cE_R +\NL\sigma_H^2\frac{\sigma_{\tilde{w}}^2}{\sigma_w^2}\cdot c(\cE_R)} \\
{\rm s.t}~ &\max \{0,\mu,\bPave(\tau_R+\tau_F)-\bPt \tau_F\} \leq \cE_R \leq \notag \\
&\min \{\bPL \tau_R,\ \bPave(\tau_R+\tau_F)-\tgam\}, \notag
\end{align}
where
\begin{equation}
\alpha(\cE_R)=\frac{\bar{P}_ave(\tau_R+\tau_F)-\tgam-\cE_R}{\tau_F+\sigma_G^2\tgam/\sigma_v^2},
\end{equation}
and
\begin{equation}
\cE_F(\cE_R)=\tgam\left(\frac{\sigma_G^2}{\sigma_v^2}\cdot \alpha(\cE_R)+1\right).
\end{equation} The optimal $\cE_F$ and $\sigma_a^2$ are given by
$\cE_F^\star=\cE_F(\cE_R^\star)$ and $(\sa2)^\star=\frac{\alpha(\cE_R^\star)}{(\Nt-\NL)}$.
\end{Prop}
Proposition 1 implies that the solutions of problem
\eqref{opt.prob.I1} can be efficiently obtained by simple line
search over a finite interval, when the condition in
\eqref{condition u} is fulfilled; otherwise, one can obtain a
simple closed-form solution of $\cE_R^*=0$, $\cE_F^*=\tgam$ and $(\sa2)^*=0$.

\chapter{Two-Way Training Strategy for Non-reciprocal Channels}\label{ch5.NonRecip}

{In this chapter, we consider the case of non-reciprocal channels, where the downlink and uplink channel matrices are asymmetric. In this case, the downlink channel cannot be directly inferred from the uplink channel. Therefore, an additional training stage using an echoed signal (from transmitter to LR and back to the transmitter) is needed in order to obtain an estimate of the combined downlink and uplink channel. This additional stage is referred to as round-trip training.} The proposed two-way training method for DCE in non-reciprocal case is detailed below.

{\bf Round-trip Training}: In round-trip training, the transmitter first broadcasts a random
signal then the LR will echo its received signal back. By the round-trip
procedure, the echoed signal obtained at the transmitter contains a combined
term of uplink channel and downlink channel. Then with the help of the following reverse training, the transmitter can obtain the downlink channel estimate. Specifically, the random signal sent by the transmitter is given by
\begin{equation}
\bX_{t0}=\sqrt{\frac{\cP_0 \tau_0}{\Nt}}\bC_{t0},
\end{equation}
where $\bC_{t0} \in \bbC^{\tau_0 \times \Nt}$ is the pilot matrix satisfying
$\text{Tr}(\bC_{t0}^H\bC_{t0})=\Nt$, and $\cP_0$ and $\tau_0$ represent the
training power and training length, respectively. For ease of use later,
we define the round-trip training energy as $\cE_0\triangleq \cP_0 \tau_0$.
The received signal at the LR is given by
\begin{equation}
\bY_{L0}=\bX_{t0}\bHd+\bW_0,
\end{equation}
where each element of $\bHd$ is assumed to be $\iid$ complex Gaussian random variable with zero mean and variance
equal to $\sHd2$ and $\bW_0\in \bbC^{\tau_0\times \NL}$ is the additive white
Gaussian noise (AWGN) matrix with each
entry having zero mean and variance $\sw2$. Then the LR amplifies and forwards
its received signal back to the transmitter. The echoed signal at the transmitter
is given by
\begin{align}\label{eq.echoed.signal}
\bY_{t1}&=\alpha\bY_{L0}\bHu+\bWt_1\\
     &=\alpha\bX_{t0}\bHd\bHu+\alpha\bW_0\bHu+\bWt_1 \notag
\end{align}
where each element of $\bHu$ is assumed to be $\iid$ complex Gaussian random
variables with zero mean and variance $\sHu2$, $\bWt_1\in \bbC^{\tau_0 \times \Nt}$
is the AWGN matrix at the transmitter with the power of each entry equal to $\swt2$.
The amplifying gain at the LR is given by
\begin{align}
\alpha&=\sqrt{\frac{\cP_1\tau_0}{\cP_0 \tau_0\NL\sHd2+\tau_0 \NL \sw2}}\notag \\
&=\sqrt{\frac{\cE_1}{\cE_0\NL\sHd2+\tau_0\NL\sw2}}
\end{align}
where $\cP_1$ is the transmission power and $\cE_1\triangleq \cP_1 \tau_0$ is the energy
of the transmitted symbol.
Since the random signal $\bX_{t0}$ is available at
the transmitter, it is able to
obtain the downlink channel estimate with a given uplink channel.
We will see how a reverse training helps the transmitter to
extract the knowledge of downlink channel $\bHd$. Note that the random signal $\bX_{t0}$
is unknown to both LR and UR, therefore the UR can not benefit from the round-trip training.

{\bf Reverse Training}: In reverse training, the LR sends a training signal
$\bX_{L2}\in\bbC^{\tau_2 \times \NL}$ to enable the uplink channel estimation at
the transmitter. Specifically, the reverse training signal is given by
\begin{equation}
\bX_{L2}=\sqrt{\frac{\cP_2\tau_2}{\NL}}\bC_{L2},
\end{equation}
where $\bC_{L2}$ is the pilot matrix which satisfies $\text{Tr}(\bC_{L2}^H\bC_{L2})=\NL$,
and $\cP_2$ and $\tau_2$ is the transmission power and training interval of the LR.
For simplicity, We define the reverse training energy as $\cE_2\triangleq \cP_2\tau_2$.
The received signal at the transmitter is given by
\begin{equation}
\bY_{t2}=\bX_{L2}\bHu+\bWt_2
\end{equation}
where $\bWt_2\in \bbC^{\tau_2\times \Nt}$ is the additive white noise matrix with each entry
having zero mean and variance $\swt2$. As the uplink channel estimate is given, the downlink
channel estimate can be acquired from the echoed signal.

{\bf Forward Training}: In forward training, the transmitter
sends AN along with the training signal to discriminate the channel estimation performances
between LR and UR. The specific description is stated in the Step II of Chapter \ref{ch2.system}. Note that we replace the
subscript by $3$ for notation consistency in this chapter, therefore
the received signals at the LR and UR are replaced by
\begin{align}\label{eq.receive.signal.LR3}
\bY_{L3}&=\sqrt{\frac{\cE_3}{\Nt}}\bC_{t3}\bHd+\bA\bNHdH\bHd+\bW_3\\
\bY_{U3}&=\sqrt{\frac{\cE_3}{\Nt}}\bC_{t3}\bG+\bA\bNHdH\bG+\bV_3
\label{eq.receive.signal.UR3}
\end{align}
and the forward training length $\tau_3$ is to substitute $\tau_F$.

Due to the complicated nature of the two-way training, finding the optimal pilot
structures may be a difficult task, which could also be different for different
objective functions, $\eg$ channel estimation error, bit error rate or ergodic
capacity, $\etc$. The practical intuition in choosing the pilot structure is 1)
to reduce the channel estimation error and 2) to reduce the transmission overhead.
In conventional channel estimation, the orthogonal structure was usually found
to be good. Note that it may not be the optimal choice for the system we are
considering. By utilizing the orthogonal training signal, the performance of
channel estimation is now determined by the training energy of each phase and
one can keep the training length minimum if the training energy can be designed
to reduce the channel estimation error. Besides, it is preferred to keep the
training length small in the secrecy channel estimation according to the remark
stated in Chapter \ref{ch4.Recip.Anal}. Hence, in this work, we choose the minimum
training length to be the number of transmit antenna, $\ie$ $\tau_0=\tau_3=\Nt$
and $\tau_2=\NL$. And we assume the unitary pilot data are used, that is
$\bC^H_{t0}\bC_{t0}=\bC_{t0}\bC^H_{t0}=\bI_{\Nt}$,
$\bC_{L2}^H\bC_{L2}=\bC_{L2}\bC^H_{L2}=\bI_{\NL}$
and $\bC_{t3}^H\bC_{t3}=\bC_{t3}\bC^H_{t3}=\bI_{\Nt}$.

\chapter{Optimal Power Allocation for DCE in Nonreciprocal Channels}\label{ch6.Nonrecip.Anal}

In this chapter, we show how the transmitter can compute the downlink channel
estimate from the training signals and analyze the channel estimation performance at both LR and UR. We
assume that the transmitter, LR and UR all employ the linear minimum mean square error
(LMMSE) criterion for channel estimation \cite{Est_Theory}. Then, we examine the optimal power allocation between training and AN signals in this case and propose an efficient solution for this problem.

\section{Channel Estimation Performance at Transmitter}

In this section , we show how the transmitter computes the downlink channel estimate from the
reverse and round-trip training signals. Specifically, with the help of reverse training and  by employing the LMMSE estimator, the estimate of the uplink channel $\bHu$ can first be computed as \cite{Est_Theory}
\begin{equation}
\hbHu=\sHu2\bX^H_{L2}(\sHu2\bX_{L2}\bX^H_{L2}+\swt2\bI_{\Nt})^{-1}\bY_{t2}\triangleq\bHu+\Delta\bHu
\end{equation}
where $\Delta\bHu\in\bbC^{\NL\times\Nt}$ is the estimation error matrix with
correlation matrix given by
\begin{equation}\label{eq.mse.uplinkbHu}
\text{E}\{ (\Delta\bHu)^H\Delta\bHu\}=\NL\left( \frac{1}{\sHu2}+\frac{\cE_2}{\NL\swt2}\right)^{-1}\bI_{\Nt}.
\end{equation}
With the uplink channel estimate $\hbHu$ being available at the transmitter, we can
rewrite the echoed signal (\ref{eq.echoed.signal}) as
\begin{align}\label{eq.echoed.signal.givenhbHu}
\bY_{t1}&=\alpha \bX_{t0}\bHd(\hbHu-\Delta\bHu)+\alpha\bW_0(\hbHu-\Delta\bHu)+\bWt_1\\
&=\alpha \bX_{t0}\bHd\hbHu+(-\alpha\bW_0\hbHu-\alpha\bX_{t0}\bHd\Delta\bHu+\alpha\bW_0\Delta\bHu+\bWt_1)\notag.
\end{align}
To employ the LMMSE criterion for the downlink channel estimation at the transmitter,
it is easier to empress (\ref{eq.echoed.signal.givenhbHu}) in the vector form as
\begin{equation}
\mathbf{y}_{t1}=\alpha(\hbHu^T\otimes \bX_{t0})\mathbf{h}_d-\alpha(\Delta\bHu^T\otimes \bX_{t0})\mathbf{h}_d
+\alpha(\hbHu^T\otimes \bI_{\Nt})\mathbf{w}_0-\alpha(\Delta\bHu^T\otimes \bI_{\Nt})\mathbf{w}_0
+\mathbf{\tilde{w}}_1
\end{equation}
where the fact that $\text{vec}(ABC)=(C^T\otimes A)\text{vec}(B)$ is used,
$\mathbf{y}_{t1}=\text{vec}(\bY_{t1})$ is formed by stacking the columns of $\bY_{t1}$ and
so do $\mathbf{h}_{d}=\text{vec}(\bHd)$, $\mathbf{w}_{0}=\text{vec}(\bW_0)$, and $\mathbf{\tilde{w}}_1=\text{vec}(\bWt_1)$.
As $\hbHu$ is given at the transmitter, by the fact that
$\hbHu$ and $\Delta\bHu$ are uncorrelated due to the orthogonality principle \cite{Est_Theory},
the premise of $\bC_{t0}\bC^H_{t0}=\bC^H_{t0}\bC_{t0}=\bI_{\Nt}$ and (\ref{eq.mse.uplinkbHu}), the LMMSE estimate
of downlink channel $\mathbf{h}_d$ and thus its matrix form
are respectively given by
\begin{align}
\mathbf{\hat{h}}_{d,t}
&=\frac{1}{\alpha\sw2}\left(\frac{1}{\sHd2}+\frac{\cE_0}{\Nt\sw2}\right)^{-1}\left(\hbHu^*
\left( (\hbHu^T\hbHu^*)+\beta\bI_{\Nt}\right)^{-1}
\otimes \bX_{t0}^H\right)\mathbf{y}_{t1} \\
&\triangleq\mathbf{h}_d+\Delta\mathbf{h}_{d,t}\\
\hbHdt&=\frac{1}{\alpha\sw2}\left(\frac{1}{\sHd2}+\frac{ \cE_0}{\Nt\sw2}\right)^{-1}
\bX_{t0}^H\bY_{t1}\left( (\hbHu^H\hbHu)+\beta\bI_{\Nt}\right)^{-1}\hbHu^H\\
&\triangleq\bHd+\Delta\mathbf{H}_{d,t}
\end{align}
where
\begin{equation}
\beta=
\NL\left( \frac{1}{\sHu2}+\frac{\cE_2}{\NL\swt2}\right)^{-1}+
\frac{\swt2}{\alpha^2\sHd2\sw2}\left(\frac{1}{\sHd2}+\frac{ \cE_0}{\Nt\sw2}\right)^{-1}
\end{equation}
and $\Delta\mathbf{h}_{d,t}\in\bbC^{\Nt\NL\times 1}$ is the estimation error
vector at the transmitter. The correlation matrix of $\Delta\mathbf{h}_{d,t}$
conditioned on a given $\hbHu$ is given by
\begin{equation}\label{eq.mse.downlinkHd.tx}
\text{E}\{\Delta\mathbf{h}_{d,t}(\Delta\mathbf{h}_{d,t})^H|\hbHu\}
=\left[\sHd2\bI_{\NL}-\sHd2\frac{\sHd2 \cE_0}{\sHd2 \cE_0+\Nt\sw2}\left(
\left(\frac{1}{\beta}\hbHu^*\hbHu^T\right)^{-1}+\bI_{\NL}\right)^{-1}\right]\otimes \bI_{\Nt}
\end{equation}
Note that for differentiating from the downlink channel estimate of the LR, we denote
the downlink channel estimate of the transmitter as $\hbHdt$. The matrix consisting of
the basis of left null space of $\hbHdt$ is replaced as $\bNHdt$.

\section{Channel Estimation Performance at LR and UR}
In this section we analyze the channel estimation performance of the LR and UR.
We first consider the channel estimation at the LR. Due to the fact that $\bNHdH\hbHdt=\mathbf{0}$
the received signal of LR (\ref{eq.receive.signal.LR3}) can be written as
\begin{equation}\label{eq.receive.signal.LR2}
\bY_{L3}=\bar{\bC}_{t3}\bHd-\bA\bNHdHt\Delta\mathbf{H}_{d,t}+\bW_3.
\end{equation}
where $\bar{\bC}_{t3}\triangleq\sqrt{\frac{\cE_3}{\Nt}}\bC_{t3}$.
To apply the LMMSE criterion for the downlink channel estimation of the LR, let us vectorize (\ref{eq.receive.signal.LR2}) as
\begin{equation}
\mathbf{y}_{L3}=\left(\bI_{\NL}\otimes \bar{\bC}_{t3}\right)\mathbf{h}_d
-(\bI_{\NL}\otimes \bA\bNHdH)\Delta\mathbf{h}_{d,t}+\mathbf{w}_3
\end{equation}
where $\mathbf{h}_{d,t}=\text{vec}(\bHd)$ and $\mathbf{w}_3=\text{vec}(\bW_3)$.
Then the channel estimate of $\mathbf{h}_d$ is given by
\begin{equation}
\mathbf{\hat{h}}_d=\bC_{h_dy_{L3}}\bC^{-1}_{y_{L3}y_{L3}}\mathbf{y}_{L3}
\end{equation}
where
\begin{equation}\label{eq.covar.hdyL3}
\bC_{h_dy_{L3}}=\text{E}\{\mathbf{h}_d\mathbf{y}_{L3}^H\}=\sHd2\left(\bI_{\NL}\otimes \bar{\bC}_{t3}\right)
\end{equation}
is the covariance matrix between $\mathbf{h}_d$ and $\mathbf{y}_{L3}$ and
\begin{align}
\bC_{y_{L3}y_{L3}}&=\text{E}\{\mathbf{y}_{L3}\mathbf{y}_{L3}^H\}\\
&=\sHd2\left(\bI_{\NL}\otimes \bar{\bC}_{t3}\bar{\bC}_{t3}^H\right) \label{eq.covar.yL3}
+\text{E}\{(\bI_{\NL}\otimes \bA\bNHdH)\Delta\mathbf{h}_{d,t}
\Delta\mathbf{h}^H_{d,t}(\bI_{\NL}\otimes \bA\bNHdH)^H\}+\sw2(\bI_{\NL}\otimes \bI_{\Nt})
\end{align}
is the covariance matrix of $\mathbf{y}_{L3}$. The expectation in (\ref{eq.covar.yL3})
is taken over all the random variables including $\bA$, $\Delta\mathbf{h}_{d,t}$ and
$\hbHdt$ of which the last two are functions of the random matrix $\hbHu$.
With the law of iterated expectations, $\ie$ $\text{E}\{X\}=\text{E}\{\text{E}\{X|Y\}\}$,
the second term of (\ref{eq.covar.yL3}) can be written as
\begin{equation}\label{eq.covar.yL3.2term}
\text{E}_{\hbHu}\{ \text{E}_{\bA,\hbHdt}\{(\bI_{\NL}\otimes \bA\bNHdHt)
\text{E}\{\Delta\mathbf{h}_{d,t}\Delta\mathbf{h}^H_{d,t}|\hbHdt,\hbHu\}
(\bI_{\NL}\otimes \bA\bNHdH)^H|\hbHu  \}\}
\end{equation}
where a fact that the random matrix $\bA$ is independent of $\Delta\mathbf{h}_{d,t}$
is used. Since the term $\text{E}\{\Delta\mathbf{h}_{d,t}\Delta\mathbf{h}^H_{d,t}|\hbHdt,\hbHu\}$
is not easy to tackle, we made an assumption that $\hbHdt$ is Gaussian distributed
under a given $\hbHu$. In this case, $\Delta \bHdt=\bHd-\hbHdt$ is also Gaussian
distributed and so does its vector form $\Delta \mathbf{h}_{d,t}$. We know that
$\Delta \bHdt$ is uncorrelated to $\hbHdt$ refering to the orthogonality principle,
therefore $\Delta \mathbf{h}_{d,t}$ and $\hbHdt$ are independent due to our
imposed Gaussian assumption. The equation in (\ref{eq.covar.yL3.2term}) is then given by
\begin{equation}\label{eq.covar.yL3.2term.2}
\text{E}_{\hbHu}\{ \text{E}_{\bA,\hbHdt}\{(\bI_{\NL}\otimes \bA\bNHdHt)
\text{E}\{\Delta\mathbf{h}_{d,t}\Delta\mathbf{h}^H_{d,t}|\hbHu\}
(\bI_{\NL}\otimes \bA\bNHdH)^H|\hbHu  \}\}
\end{equation}
Substituting (\ref{eq.mse.downlinkHd.tx}) and the fact that $\bNHdHt\bNHdt=\bI_{\Nt-\NL}$ into
(\ref{eq.covar.yL3.2term.2}), we obtain
\begin{align}\label{eq.covar.yL3.2term.3}
(\Nt-\NL)\sa2\left[\sHd2\bI_{\NL}-\sHd2\frac{\sHd2 \cE_0}{\sHd2 \cE_0+\sw2}\text{E}\Biggl\{\left(
\left(\frac{1}{\beta}\hbHu^*\hbHu^T\right)^{-1}+\bI_{\NL}\right)^{-1}\Biggr\}\right]\otimes \bI_{\Nt}
\end{align}
The Hermitian term $\hbHu\hbHu^H$ can be factorized into
\begin{equation}
\hbHu\hbHu^H=\mathbf{U}\mathbf{\Lambda}\mathbf{U}^H
\end{equation}
where $\mathbf{U}\in \bbC^{\NL\times \NL}$ is the matrix whose columns are consisting
of eigenvectors of $\hbHu\hbHu^H$ and $\mathbf{\Lambda}=\text{diag}(\lambda_1,\dots,\lambda_{\NL})$
is the diagonal matrix with diagonal elements being nonzero and unordered eigenvalues
of $\hbHu\hbHu^H$. Since the elements of both the uplink channel $\bHu$ and the noise
matrix $\bWt_2$ are $\iid$ Gaussian distributed and the reverse training $\bX_{L2}$
is assumed to be semi-unitary, each entry of $\hbHu$ is $\iid$ Gaussian distributed.
Referring to \cite{Matrix_variable_dist}, we know that $\hbHu\hbHu^H$ has a Wishart
distribution with $\Nt$ degrees of freedom and its mean is given by
\begin{equation}
\text{E}\{\hbHu\hbHu^H\}=\Nt \frac{\sigma_{H_u}^4 \cE_2}{\sHu2 \cE_2+\NL\swt2}\bI_{\NL}
\triangleq\Nt \sigma^2\bI_{\NL}\notag
\end{equation}
where
\begin{equation}\label{eq.sigma}
\sigma^2\triangleq \frac{\sigma_{H_u}^4 \cE_2}{\sHu2 \cE_2+\NL\swt2}
\end{equation}
is the variance of each $\iid$ random variable of $\hbHu$.
Both the density function of $\hbHu\hbHu^H$ \cite{Matrix_variable_dist} and the Jacobian
of the eigenvalue value decomposition of $\hbHu\hbHu^H$ \cite{Random_matrix_theo} can be divided into the product
of functions of $\mathbf{\Lambda}$ and $\mathbf{U}$, thus we conclude
that $\mathbf{\Lambda}$ and $\mathbf{U}$ are independent.
With the independency and applying the law of iterated expectations, the equation in (\ref{eq.covar.yL3.2term.3}) becomes
\begin{align} \label{eq.covar.yL3.2term.4}
&(\Nt-\NL)\sa2\left[\sHd2\bI_{\NL}-\sHd2\frac{\sHd2 \cE_0}{\sHd2 \cE_0+\Nt\sw2}
\text{E}_{\mathbf{U}}\Biggl\{\mathbf{U}\cdot
\text{E}_{\mathbf{\Lambda}}\biggl\{\left(\beta\mathbf{\Lambda}^{-1}+\bI_{\NL}\right)^{-1}\biggr\}
\mathbf{U}^H\Biggr\}\right]\otimes \bI_{\Nt}\\
=&(\Nt-\NL)\sa2\left[\sHd2-\sHd2\frac{\sHd2 \cE_0}{\sHd2 \cE_0+\Nt\sw2}\label{eq.covar.yL3.2term.5}
\text{E}_{\lambda_1}\biggl\{\left(\frac{1}{\beta/\lambda_1+1}\right)\biggr\}
\right]\bI_{\NL}\otimes \bI_{\Nt}
\end{align}
where the equality holds since the eigenvalues of the Wishart distributed matrix $\hbHu\hbHu^H$
have identical distributions as any one of the unordered eigenvalues \cite{Unordered_eigenvalue_dist}.
Replacing (\ref{eq.covar.yL3.2term.5}) in (\ref{eq.covar.yL3}),
we have an approximation of the covariance matrix of $\mathbf{y}_{L3}$ as
\begin{align}
\nonumber&\bC_{y_{L3}y_{L3}}\approx\\
& \bI_{\NL}\!\otimes\! \Biggl\{\!\sHd2\bar{\bC}_{t3}\bar{\bC}_{t3}^H
+\left[(\Nt-\NL)\sa2\left(\sHd2-\sHd2\frac{\sHd2 \cE_0}{\sHd2 \cE_0+\Nt\sw2}
\text{E}_{\lambda_1}\biggl\{\left(\frac{1}{\beta/\lambda_1+1}\right)\biggr\}\right)+\sw2\right]\bI_{\Nt}
\!\Biggr\}\label{eq.covar.yL3.2}
\end{align}
The normalized mean squared error (NMSE) of $\hbHd$ can be computed as
\begin{align}
\text{NMSE}_L&=\frac{\text{Tr}(\text{E}\{\Delta\mathbf{h}_d\Delta\mathbf{h}^H_d\})}{\Nt\NL}\\
&=\frac{\text{Tr}\left(\sHd2\bI_{\NL\Nt}-\bC_{h_dy_{L3}}\bC^{-1}_{y_{L3}y_{L3}}\bC^H_{h_dy_{L3}}\right)}{\Nt\NL}
\label{eq.NMSEL.def}
\end{align}
Substituting (\ref{eq.covar.hdyL3}) and (\ref{eq.covar.yL3.2}) into (\ref{eq.NMSEL.def}), we have
an approximation for the NMSE of the LR as
\begin{equation}\label{eq.NMSEL.appr1}
\text{NMSE}_L\approx
\left(\frac{1}{\sHd2}+\frac{\cE_3}{\Nt}\frac{1}{(\Nt-\NL)\sa2\left(\sHd2-\sHd2\frac{\sHd2 \cE_0}{\sHd2 \cE_0+\Nt\sw2}
\text{E}_{\lambda_1}\biggl\{\left(\frac{1}{\beta/\lambda_1+1}\right)\biggr\}\right)+\sw2} \right)^{-1}
\end{equation}
For $\Nt\gg 1$, the distribution of the eigenvalues of $\hbHu\hbHu^H$ is asymptotically
approximated to a Gaussian distribution \cite{Limiting_dist_eigenvalue}, that is $\lambda_1\overset{a.}{\thicksim} \mathcal{N}(\Nt\sigma^2,\Nt\sigma^4)$ where
$\sigma^2$ is given in (\ref{eq.sigma}). However, the expectation term in
(\ref{eq.NMSEL.appr1}) is intractable, we instead apply the Jensen's
inequality and take its lower bound as an approximation. Hence, we have
\begin{equation}\label{eq.NMSEL.appr2}
\text{NMSE}_L\approx
\left(\frac{1}{\sHd2}+\frac{\cE_3}{\Nt}\frac{1}{(\Nt-\NL)\sa2\left(\sHd2-\sHd2\frac{\sHd2 \cE_0}{\sHd2 \cE_0+\Nt\sw2}
\left(\frac{\Nt\sigma}{\beta+\Nt\sigma}\right)\right)+\sw2} \right)^{-1}
\end{equation}
On the other hand, the NMSE performance of the UR is analyzed as follows.
The received signal of UR (\ref{eq.receive.signal.UR3}) can be vectorized as
\begin{equation}
\mathbf{y}_{U3}=\text{vec}(\bY_{U3})=(\bI_{\NU}\otimes \bar{\bC}_{t3})\mathbf{g}
+(\bI_{\NU}\otimes \bA\bNHdH)\mathbf{g}+\mathbf{v}_3
\end{equation}
where $\bar{\bC}_{t3}=\sqrt{\frac{\cE_3}{\Nt}}\bC_{t3}$, $\mathbf{g}=\text{vec}(\bG)$
and $\mathbf{v}_3=\text{vec}(\bV_3)$. The covariance matrix between $\mathbf{g}$
and $\mathbf{y}_{U3}$ and the covariance matrix of $\mathbf{y}_{U3}$ are respectively given by
\begin{align}
\bC_{g,y_{U3}}&=\sG2(\bI_{\NU}\otimes \bar{\bC}^H_{t3})\\
\bC_{y_{U3}y_{U3}}&=\bI_{\NU}\otimes \left[\sG2\bar{\bC}_{t3}\bar{\bC}^H_{t3}+
\left(\sG2(\Nt-\NL)\sa2+\sv2 \right)\bI_{\Nt} \right]
\end{align}
Hence, the NMSE of the UR is given by
\begin{align}
\text{NMSE}_U&=\frac{\text{Tr}(\sG2\bI_{\Nt\NU}-
\bC_{g,y_{U3}}\bC^{-1}_{y_{U3}y_{U3}}\bC^H_{g,y_{U3}})}{\Nt\NU}\notag\\
&=\frac{\text{Tr}\left(\bI_{\NU}\otimes \Biggl\{\sG2\bI_{\Nt}-\sG2\bar{\bC}_{t3}\left[\sG2\bar{\bC}_{t3}\bar{\bC}^H_{t3}+
\left(\sG2(\Nt-\NL)\sa2+\sv2 \right)\bI_{\Nt} \right]^{-1}\sG2\bar{\bC}_{t3}\Biggr\}\right)}{\Nt\NU}\notag\\
&=\left( \frac{1}{\sG2}+\frac{\cE_3}{\Nt}\frac{1}{\sG2(\Nt-\NL)\sa2+\sv2}\right)^{-1}\label{eq.NMSEU}
\end{align}

\section{Optimal Power Allocation between Training and AN Signals}
With (\ref{eq.NMSEL.appr2}) and (\ref{eq.NMSEU}), we can jointly design the power values
of $\{\cE_0,~\cE_1,~\cE_2,~\cE_3,~\sa2\}$ by considering the following power allocation problem
\begin{subequations}\label{prob.Nonrecip}
\begin{align}
\underset{\cE_0,\cE_1,\cE_2,\cE_3,\sa2\geq 0}{\min}\ \ &\text{NMSE}_L \\
\text{s.t.}\ \ \ \ \ \ \ \ &\text{NMSE}_U\geq \gamma\\
&\cE_0+\cE_1+\cE_2+\cE_3+(\Nt-\NL)\sa2 \Nt\leq P_{ave}(\Nt+\Nt+\NL+\Nt) \\
&\cE_0+\cE_3+(\Nt-\NL)\sa2 \Nt\leq \bar{P}_t(\Nt+\Nt)\\
&\cE_1+\cE_2\leq \bar{P}_L(\Nt+\NL).
\end{align}
\end{subequations}
Here, we aim to minimize the NMSE of LR subject to the constraint that the NMSE of UR remains above a preset
lower limit $\gamma$. We also consider the average power constraint $P_{ave}$ and two individual power
constraints $\bar{P}_t$ and $\bar{P}_L$ at the transmitter and LR, respectively. However, the problem is not easily solvable. To obtain an efficient solution, we resort to the monomial
approximation and the condensation method often adopted in the field of geometric programming (GP)
\cite{Tutorial_GP}. Details are given in the Appendix \ref{apped.GPconden}.

\chapter{Numerical Results and Discussions}\label{ch7.simulation}

\begin{figure}[t]
\begin{center}
\includegraphics[width=5.3in] {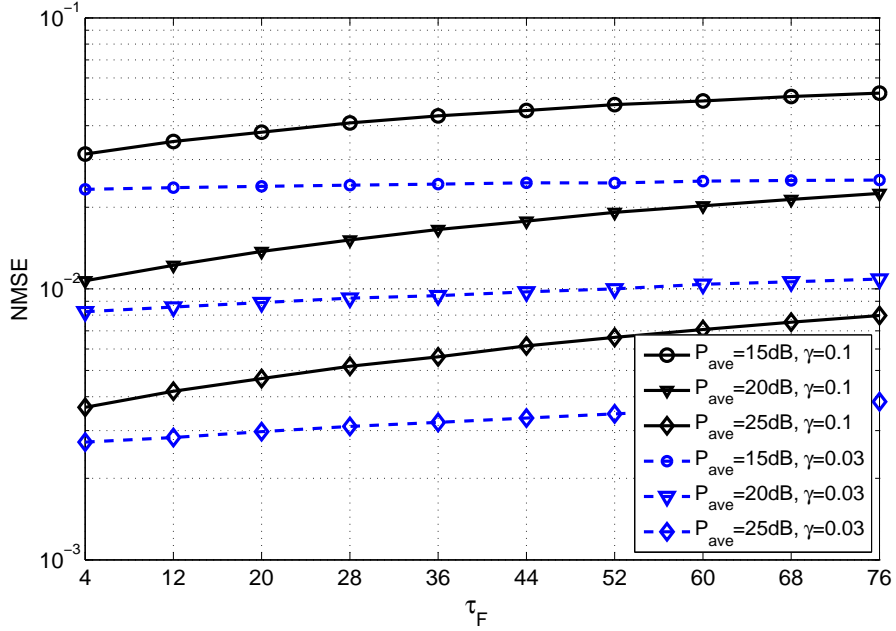}
\caption{NMSE performance versus forward training $\tau_F$ for $\bar{P}_t=30$ dB and
$\bar{P}_L=20$ dB.}
\label{fig-rec_NMSEvsTF}
\end{center}
\vspace{-0.25in}
\end{figure}

In this chapter, we present numerical results to demonstrate the effectiveness of the proposed DCE schemes. We
consider the MIMO wireless system as described in Chapter
\ref{ch2.system} with $\Nt=4$, $\NL=2$ and $\NU=2$. The elements of
the channel matrices $\bH$ and $\bG$ are $\iid$ complex Gaussian
distributed with zero mean and unit variance
($\sigma_H^2=\sigma_G^2=1$). Each entry of additive white noise
matrices $\bWt$, $\bW$ and $\bV$ is also $\iid$  complex Gaussian
distributed with zero mean and unit variance, $\ie$
$\sigma_{\tilde{w}}^2=\sigma_w^2=\sigma_v^2=1$. Moreover, the
training lengths are set to be the antenna number of the terminal
which transmits that training signal, $\ie$ $\tau_R=\NL=2$ and
$\tau_F=\Nt=4$ for the reciprocal case\footnote{In
pure channel estimation, it is preferred to keep the
training length minimum in uncorrelated channel and white noise
\cite{Hassibi_howmuchtraining}. We show in Figure \ref{fig-rec_NMSEvsTF}
that the training length is better to choose as smallest length
$\ie$ the number of transmit antenna in the secrecy channel estimation scheme.}
and $\tau_0=\tau_3=\Nt=4$ and $\tau_2=\NL=2$ for the non-reciprocal case.
Note that the overall training time is larger than
the sum of all training length due to the processing time at the
transmitter.
Besides, the individual power constraints of the transmitter and the LR are
respectively assigned as $\bPt=30$ dB and $\bPL=20$ dB.
We incorporate an NMSE lower bound for comparison. The
lower bound for reciprocal and non-reciprocal case are respectively
given by
\begin{equation}
\mbox{NMSE}_{\mbox{LB},rec}=\left( \frac{1}{\sigma_H^2}+\frac{\min\{\bPt\Nt ,\ \bPave(\NL+\Nt)\}}{\Nt\sigma_w^2}\right)^{-1}\!\!\!\!
\end{equation}
and
\begin{equation}
\mbox{NMSE}_{\mbox{LB},nonrec}=\left( \frac{1}{\sigma_H^2}+\frac{\min\{2\bPt\Nt,\ \bPave(3\Nt+\NL)\}}{\Nt\sigma_w^2}\right)^{-1}\!\!\!\!
\end{equation}
which both stand for the minimum achievable NMSE at the LR when
$\sigma_a^2=0$, $\ie$ no AN exists.

Figure \ref{fig-rec_NMSEvsTF} shows the NMSE performance of LR versus
the forward training length $\tau_F$ under the constant energy constraints.
In the reciprocal case,
the average energy constraint is given by $\bPave(\Nt+\NL)$
and the individual energy constraints of the transmitter and LR are
respectively given by $\bPt\Nt$ and $\bPL\NL$.
We compare different average power constraints $\bPave=15$ dB,
$20$ dB and $25$ dB and different lower limit values $\gamma=0.1$ and $0.03$.
We see from Fig. \ref{fig-rec_NMSEvsTF} that the NMSE value of LR is
monotonically non-decreasing with respect to the training length $\tau_F$.
It shows that in secrecy channel estimation it is better to keep
the training length as small as possible.
This is due to the fact that as the forward training length increases,
it takes more AN energy to satisfy the lower limit constraint on the UR
thus the budget for the training energies is sacrificed.
Moreover, we see that the lines of $\gamma=0.1$ are more steeper than
those of $\gamma=0.03$. The trade-off between the AN energy and training
energies is more explicit as the lower limit is severer.

\begin{figure}[t]
\begin{center}
\includegraphics[width=5.2in] {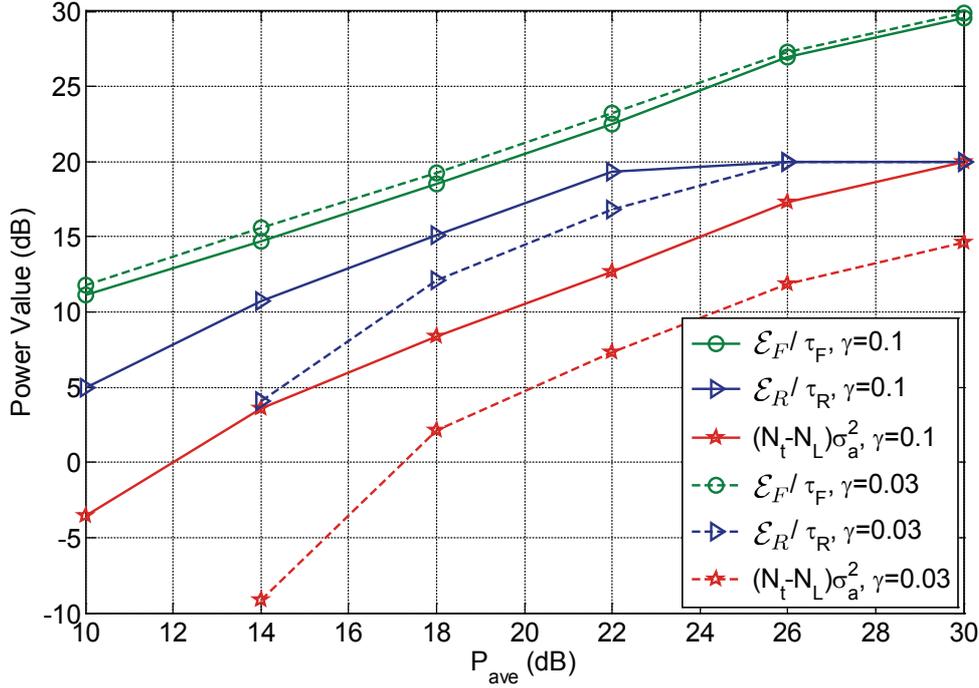}
\caption{Power allocation among reverse and forward training powers
$\cE_R/\tau_R$, $\cE_F/\tau_F$ and AN power $(\Nt-\NL)\sigma_a^2$.}
\label{fig-PowAllo_r01003}
\end{center}
\vspace{-0.35in}
\end{figure}
\begin{figure}[t]
\begin{center}
\includegraphics[width=5.2in] {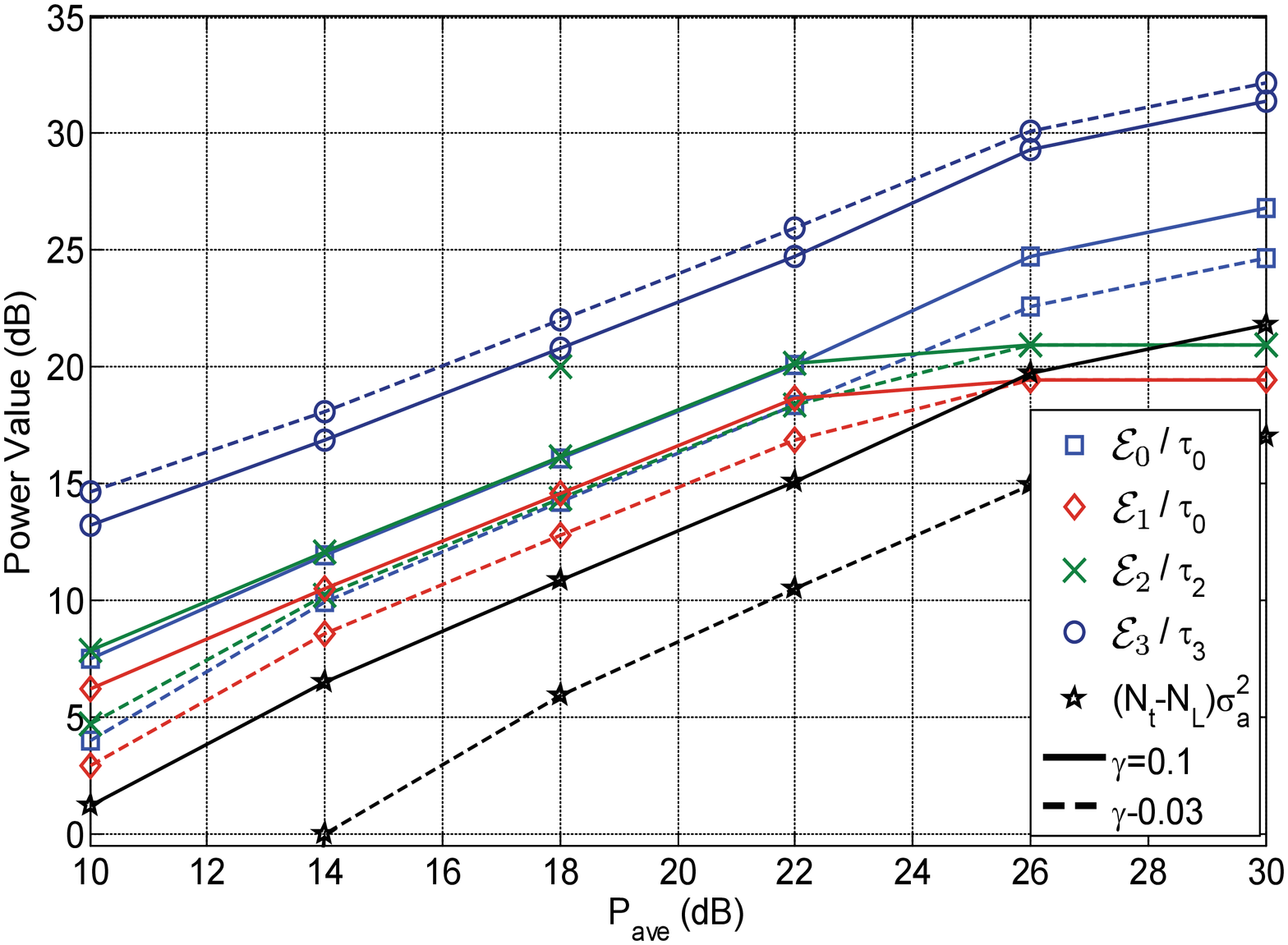}
\caption{Power allocation among the training powers $\cE_0/\Nt$, $\cE_1/\Nt$ $\cE_2/\NL$ $\cE_3/\Nt$ and AN power $(\Nt-\NL)\sigma_a^2$.}
\label{fig-PowAllononrecip}
\end{center}
\vspace{-0.35in}
\end{figure}

Figure \ref{fig-PowAllo_r01003} shows the optimal allocation of the
reciprocal case among the reverse and forward training powers
$\cE_R/\tau_R$, $\cE_F/\tau_F$ and the AN power
$(\Nt-\NL)\sigma_a^2$ versus average power constraint $\bPave$. We
compare two different lower limit values $\gamma=0.1$ and
$\gamma=0.03$. We see from Fig. \ref{fig-PowAllo_r01003}
that it is desirable to allocate more power to the AN and less power to the
forward training as $\gamma$ increase from $0.03$ to $0.1$. This is
due to the fact that the forward training signal benefits the LR and
the UR equally while the AN primarily degrades the UR's estimation
performance. In addition, we see that the reverse training power
increases with $\gamma$, since the reverse training power
mainly determines the subspace into which the AN is transmitted,
which helps to reduce the interference caused by the AN on the LR.
Note that when $\bPave=10$ dB and $\gamma=0.03$, this is the
case with $\gamma$ out of the interesting interval (\ref{cond.gamma}),
the reverse training power and AN power both equal to 0
which can not be showed in the log-value.

On the other hand,
Figure \ref{fig-PowAllononrecip} shows the power allocation of the
non-reciprocal case among the round-trip training powers
$\cE_0/\Nt$ and $\cE_1/\Nt$, reverse and forward training powers
$\cE_2/\NL$ and $\cE_3/\NL$ and the AN power $(\Nt-\NL)\sa2$ versus
average power constraint $\bPave$. We have similar observation about
the allocation between the forward training power and AN power to that
of the reciprocal case. We see from Fig. \ref{fig-PowAllononrecip}
that the round-trip and reverse training powers all increase with
respect to $\gamma$, since these powers play the role to design the
placement of AN for minimizing the interference on the LR.
As $\gamma$ increases, so does the AN power, there needs more
round-trip and reverse training powers to decrease the damage
cause by the AN on the LR.

\begin{figure}[t]
\begin{center}
\subfigure[Reciprocal case]{
\label{fig-NMSE_r39}
\includegraphics[width=5.3in]{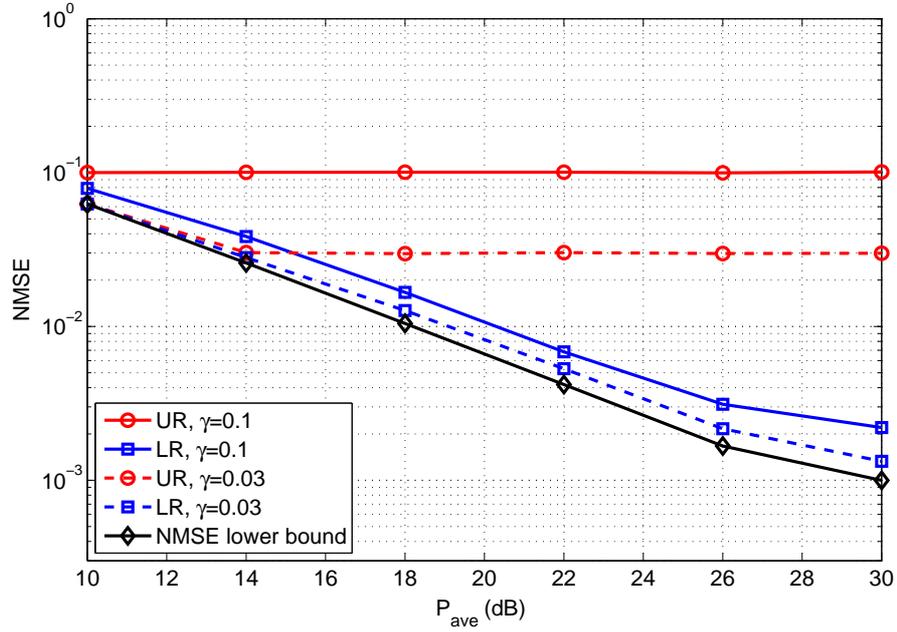}}
\subfigure[Non-reciprocal case]{
\label{fig-NMSE_nonrecip}
\includegraphics[width=5.3in]{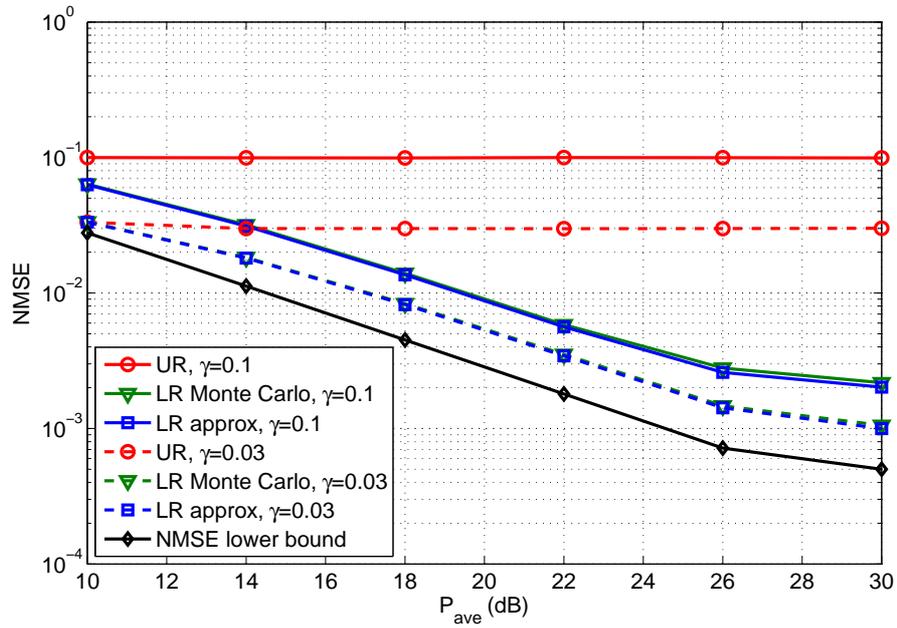}}
\caption{NMSE performance of the proposed DCE scheme for the reciprocal and non-reciprocal case.}
\label{fig-NMSE}
\end{center}
\vspace{-0.25in}
\end{figure}

Figure \ref{fig-NMSE_r39} and Figure \ref{fig-NMSE_nonrecip} show
the NMSE performance of the LR and UR versus average power constraint
$\bPave$ respectively for the reciprocal and the non-reciprocal channel.
We compare two different lower limit values $\gamma=0.1$ and
$\gamma=0.03$ in both figures. From Fig. \ref{fig-NMSE},
we observed that the NMSE of the UR meets the lower limit in
both reciprocal and non-reciprocal case. Furthermore, the proposed
DCE scheme constrains the UR's NMSE well above $\gamma$.
In addition, from Fig. \ref{fig-NMSE_nonrecip}, we see that the
approximation of LR's NMSE (\ref{eq.NMSEL.appr2})
is quite close to the Monte-Carlo simulation result of LR's NMSE.

\begin{figure}[t]
\begin{center}
\subfigure[Reciprocal case: 64-QAM OSTBC]{
\label{fig-SER_rec_64}
\includegraphics[width=5.3in]{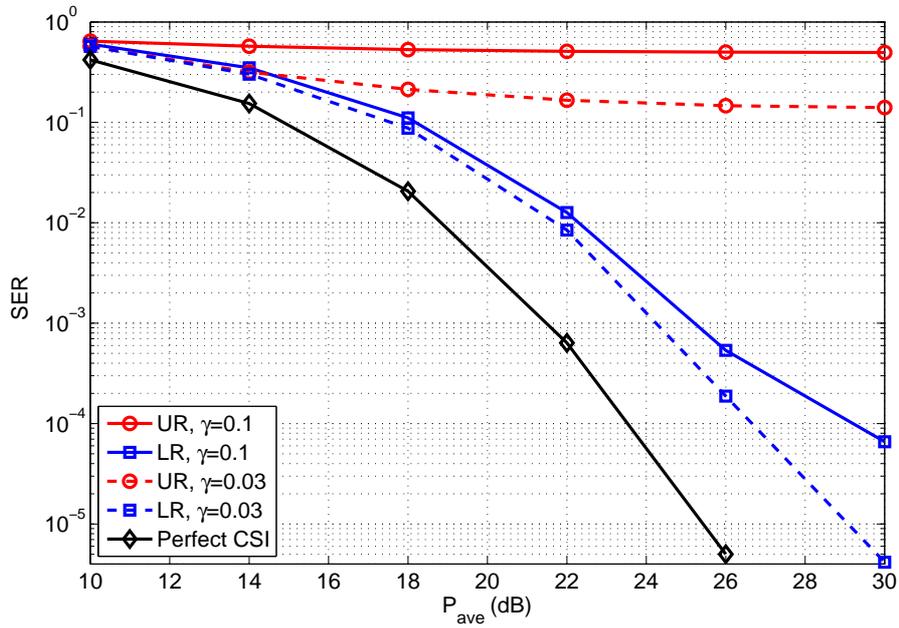}}
\subfigure[Non-reciprocal case: 64-QAM OSTBC]{
\label{fig-SER_nonrecip_64}
\includegraphics[width=5.3in]{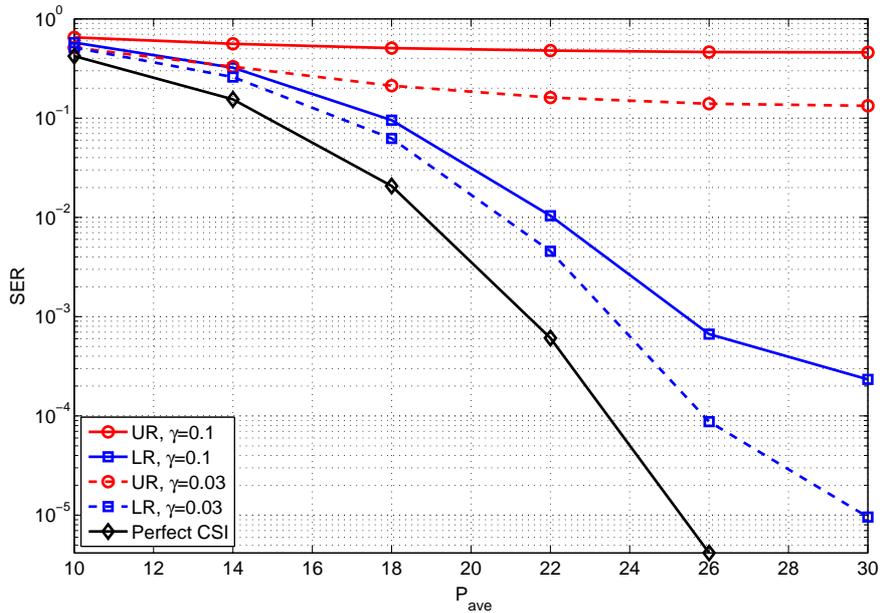}}
\vfill
\caption{SER performance of the LR and UR in an OSTBC system with the channel estimates obtained by the proposed DCE scheme.}
\label{fig-SER}
\end{center}
\vspace{-0.3in}
\end{figure}

In Figure \ref{fig-SER}, we show the symbol error rate (SER) at LR
and UR versus the average power constraint $\bPave$ in the data
transmission phase. We consider the scenario where the transmitter
sends a $4\times 4$ complex orthogonal STBC (OSTBC) with $\Nt=4$.
The code length is equal to four and each code block contains three QAM source symbols \cite{ref.STBC}. The data transmission power is set
to $\bPave$. Both LR and the UR will exploit their channel
estimates obtained by the proposed DCE to decode the received
symbols. In this Monte-Carlo simulation, the SER is obtained by
averaging over $50000$ channel realization and OSTBCs.
In particular, Figure \ref{fig-SER_rec_64} presents the associated average SERs
for 64-QAM OSTBC in the reciprocal case. We see that the SER of the
LR will gradually improve while the SER of the UR remains larger than $0.1$
due to the poor channel estimation performance at the UR.
Figure \ref{fig-SER_nonrecip_64} shows the associated average SERs
for 64-QAM OSTBC in the non-reciprocal case. We have similar
observation in this case. Both figures
illustrate that, with the proposed two-way training DCE scheme, the
discrimination of the data detection performances between LR and UR can be effectively achieved. It is worthwhile to mention that
the feedback-and-retraining DCE scheme proposed in
\cite{ChangChiangHongChi_TSP2010} assumes a perfect feedback channel
with no power consumption and, thus, it is difficult to have a fair
performance comparison between the proposed scheme and that in
\cite{ChangChiangHongChi_TSP2010}.

\chapter{Conclusion}\label{ch8.conclusion}

In this thesis, we proposed a new DCE scheme based on the two-way training methodology, where both the transmitter and LR are allowed to emit training signals. In particular, training signals sent by LR are used to help the transmitter obtain an accurate estimate of the transmitter-to-LR channel.
The proposed two-way DCE scheme utilizes two phases of training in
 reciprocal channels and  three phases of training in non-reciprocal channels. The proposed training design
drastically decreases the overall training overhead compared to the original DCE scheme proposed in \cite{ChangChiangHongChi_TSP2010}. {The training and AN powers were optimized by minimizing} the NMSE of LR subject to a preset
lower limit on the NMSE of UR, an average {total power constraint, and individual power constraints over all transmitters. For the case with reciprocal
channels, the optimal power allocation  problem was reformulated into a one-variable optimization problem
which can be easily solved by simple line search. For the case with non-reciprocal channels, we derived an approximation of
LR's NMSE and utilized
monomial approximation and condensation method to obtain an approximate solution for the power allocation problem.
Numerical results were provided to
verify the efficiency of the proposed two-way DCE schemes.


\chapter{Appendix}
\section{Proof of Proposition I}
For notational simplicity, let us define $\alpha=(\Nt-\NL)\sa2$. In the following, we solve the optimization problem in two steps: (i) find the optimal values of
$\cE_F$ and $\alpha$ for any given $\cE_R$; and (ii) find the optimal value of $\cE_R$.

\noindent\underline{\bf Step i :}

 Suppose a feasible $\cE_R$ is given, the optimal values
of $\cE_F$ and $\alpha$ can be found as functions of $\cE_R$ from the
below optimization problem
\begin{subequations}\label{opt.prob.II.givenER}
\begin{align}\label{opt.prob.II1}
\max_{\cE_F,\alpha\geq 0}\ \ &\frac{(\NL\sigma_w^2+\sigma_H^2\cE_R)\cE_F}{\NL\sigma_w^2+\sigma_H^2\cdot \cE_R
+\NL\sigma_H^2\frac{\sigma_{\tilde{w}}^2}{\sigma_w^2}\cdot \alpha} \\
\mbox{s.t.}~ &\frac{\sigma_v^2\cdot \cE_F}{\sigma_G^2\cdot
\alpha+\sigma_v^2}\leq \tgam,\label{opt.prob.II2}\\
&\cE_F+\alpha\cdot \tau_F \leq \bPave(\tau_R+\tau_F)-\cE_R,\\
&\cE_F+\alpha\cdot \tau_F\leq \bPt \tau_F\label{opt.prob.II4}.
\end{align}
\end{subequations}
Note that from \eqref{opt.prob.I1} a feasible
$\cE_R$ must satisfy $\cE_R\leq\bPL \tau_R$.
In the following, we consider two different ranges of $\cE_R$.

\noindent\underline{Case 1 ($\bPave (\tau_R+\tau_F)-\tgam<\cE_R \leq \bPL \tau_R$):} If $\bPave (\tau_R+\tau_F)-\tgam< \bPL \tau_R$ holds. Since the objective function in
\eqref{opt.prob.II.givenER} is monotonically increasing with respect to $\cE_F$ but decreasing with respect to $\alpha$, by \eqref{interesting condition} and the condition of $\cE_R>\bPave (\tau_R+\tau_F)-\tgam$,
we get $\cE_F^*(\cE_R)=\bPave (\tau_R+\tau_F)-\cE_R$, $\alpha^*=0$ and hence the value of \eqref{opt.prob.II1} becomes
\begin{equation}\label{obj.inequ.agmu0}
\cE_F^*(\cE_R)=\bPave (\tau_R+\tau_F)-\cE_R,
\end{equation}
which is less than $\tgam$.

\noindent\underline{Case 2 ($\cE_R\leq \min\{\bPL
T_R,\ \bPave(\tau_R+\tau_F)-\tgam \}$):} It can be observed that if the constraint (\ref{opt.prob.II2})
is inactive we can always decrease $\alpha$ until activating the
constraint to obtain a larger objective value. If the condition
(\ref{opt.prob.II2}) is still inactive even when $\alpha=0$,
we can instead lift $\cE_F$ to achieve a larger objective value
while still satisfying \eqref{interesting case}, \eqref{interesting condition}
and the condition $\cE_R\leq \min\{\bPL
T_R,\ \bPave(\tau_R+\tau_F)-\tgam \}$.
We conclude that constraint (\ref{opt.prob.II2}) must be active at the optimum. Hence we have
\begin{equation} \label{optbc.equofa}
\cE_F^*(\cE_R)=\tgam\left( \frac{\sigma_G^2}{\sigma_v^2}\cdot \alpha^*(\cE_R)+1\right).
\end{equation}
By substituting (\ref{optbc.equofa}) into \eqref{opt.prob.II.givenER},
the problem becomes
\begin{subequations}\label{opt.prob.III.givenER}
\begin{align}\label{opt.prob.III1}
\!\!\!\!\!\!\!\!\max_{\alpha\geq 0}\
&\frac{({\sigma_G^2}/{\sigma_v^2}\cdot
\alpha+1)(\NL\swt2+\sigma_H^2\cdot
\cE_R)\tgam}{\NL\sigma_H^2\frac{\sigma_{\tilde{w}}^2}{\sigma_w^2}\cdot
\alpha+\NL\swt2+\sigma_H^2\cdot \cE_R} \\
\mbox{s.t.}\ \ \ \ &\left( \tau_F+\frac{\sigma_G^2\tgam}{\sigma_v^2}\right)\alpha+\cE_R \leq \bPave(\tau_R+\tau_F)-\tgam
\label{opt.prob.III2}\\
& \tgam \left( \frac{\sigma_G^2}{\sigma_v^2}\alpha+1\right)+\tau_F\cdot \alpha\leq \bPt \tau_F.\label{opt.prob.III3}
\end{align}\end{subequations}
The range of $\cE_R$ in this case is further divided into the following two subranges.

\begin{enumerate}
\item[(a)]
When $\cE_R < \mu$ and $\cE_R\leq \min\{\bPL
\tau_R,\ \bPave(\tau_R+\tau_F)-\tgam \}$,
where $\mu \triangleq \NL
\left(\frac{\sigma_v^2\sigma_{\tilde{w}}^2}{\sigma_G^2\sigma_w^2}-\frac{\swt2}{\sigma_H^2}\right)$,
the objective function in (\ref{opt.prob.III1}) is a
monotonically decreasing function with respect to $\alpha$. Therefore,
the optimal value of $\alpha^*(\cE_R)$ is 0 and the corresponding
optimal objective value is equal to $\tgam$.

\item[(b)] When $\mu \leq \cE_R \leq \min\{\bPL
\tau_R,\ \bPave(\tau_R+\tau_F)-\tgam \}$, the objective function in (\ref{opt.prob.III1}) is
monotonically non-decreasing with respect to $\alpha$.
For $\mu \leq \cE_R \leq \bPave(\tau_R+\tau_F)-\bPt \tau_F$,
constraint (\ref{opt.prob.III3}) must be active at the optimum
with
\begin{equation}\label{equ.c_a1}
\alpha^*=\frac{\bPt \tau_F-\tgam}{\tau_F+\sigma_G^2\tgam/\sigma_v^2}.
\end{equation}
Reversely, considering $\cE_R\geq \max\{\mu,\
\bPave(\tau_R+\tau_F)-\bPt \tau_F\}$,
the constraint (\ref{opt.prob.III2}) must be active at the optimum with
\begin{equation}\label{equ.c_a}
\alpha^*(\cE_R)=\frac{\bPave(\tau_R+\tau_F)-\tgam-\cE_R}
{\tau_F+\sigma_G^2\tgam/\sigma_v^2}
\end{equation}
Moreover, for $\cE_R\geq \mu$, the optimal objective value of
\eqref{opt.prob.III1} can be shown to be
\begin{equation}\label{obj.inequ.agmu}
\frac{\sigma_G^2c^*/\sigma_v^2+1}{\frac{\NL\sigma_H^2\sigma_{\tilde{w}}^2/\sigma_w^2}{\NL\sigma_w^2+\sigma_H^2\cdot a}c^*+1}\cdot \tgam \geq \tgam.
\end{equation}
which is no less than $\tgam$.
\end{enumerate}

\noindent\underline{\bf Step ii :}

We now solve for the optimal value of $\cE_R$. From the analysis in Step i,
a feasible $\cE_R$ satisfying $\cE_R\leq\min\{\bPL \tau_R,\ \bPave(\tau_R+\tau_F)-\tgam\}$ leads to
greater objective value than that of $\bPave(\tau_R+\tau_F)
-\tgam<\cE_R\leq \bPL \tau_R$, thus the optimal value of $\cE_R$ must lie in the former condition.
For the first case that $\mu >\min\{\bPL \tau_R,\
\bPave(\tau_R+\tau_F)-\tgam\}$, we can infer $\cE_R<\mu $ for all feasible $\cE_R$ satisfying
$\cE_R\leq\min\{\bPL \tau_R,\ \bPave(\tau_R+\tau_F)-\tgam\}$
so that $\alpha^*=0$ and thus $\cE_F^*=\tgam$. Then we get $\cE_R^*=0$ for no
need of AN. For the other case of $\mu \leq \min\{\bPL \tau_R,\
\bPave(\tau_R+\tau_F)-\tgam\}$, from (\ref{obj.inequ.agmu0}) and (\ref{obj.inequ.agmu})
we can see that the corresponding objective value for
$\max\{0,\ \mu\}\leq \cE_R
\leq \min\{\bPL \tau_R,\ \bPave(\tau_R+\tau_F)-\tgam\} $ is no less than
that for $\cE_R<\mu$. If $\mu
\leq \cE_R \leq \bPave(\tau_R+\tau_F)-\bPt \tau_F$ exists,
the optimization problem
(\ref{opt.prob.III.givenER}) becomes
\begin{align} \label{opt.prob.sub1}
\max_{\widetilde{\cE}_R\geq 0}\ \ \
&\frac{(\NL\swt2+\sigma_H^2\widetilde{\cE}_R)\cE_F^\star}{\NL\swt2+\sigma_H^2\cdot \widetilde{\cE}_R +\NL\sigma_H^2\frac{\sigma_{\tilde{w}}^2}{\sigma_w^2}\cdot \alpha^\star} \\
\mbox{s.t}\ \ \ \ \ &\max\{0,\ \mu\}\leq \widetilde{\cE}_R \leq \bPave(\tau_R+\tau_F)-\bPt \tau_F \notag
\end{align}
where $\cE_F^\star$ and $\alpha^\star$ are given by \eqref{optbc.equofa} and
\eqref{equ.c_a1} which do not depend
on $\cE_R$ in this condition. It can be observed that the objective
function (\ref{opt.prob.sub1}) is monotonically non-decreasing with
respect to $\tilde{a}$; thus the optimal value is achieved when
$\widetilde{\cE}_R^*=\bPave(\tau_R+\tau_F)-\bPt \tau_F$.
However, the corresponding
optimal objective value of (\ref{opt.prob.sub1}) is the same as the
objective value of (\ref{opt.prob.final1}) in this case. Hence, we
can have the value of $\cE_R^*$ lie in the interval $\max
\{0,\ \mu,\ \bPave(\tau_R+\tau_F)-\bPt \tau_F\} \leq \cE_R \leq \min \{\bPL
\tau_R,\ \bPave(\tau_R+\tau_F)-\tgam\} $ by solving the optimization problem
(\ref{opt.prob.final1}) and the corresponding $\alpha(\cE_R)$ and
$\cE_F(\cE_R)$ are given by (\ref{equ.c_a}) and (\ref{optbc.equofa}), respectively.

\section{Monomial Approximation and Condensation Method for the Problem in (\ref{prob.Nonrecip})}
\label{apped.GPconden}

Here, we show how to obtain an efficient solution for the problem in (\ref{prob.Nonrecip}) using monomial
approximation and condensation method. The problem in
(\ref{prob.Nonrecip}) can be stated as follows:
\begin{subequations}\label{prob.Nonrecip.2}
\begin{align}
\underset{\cE_0,\cE_1,\cE_2,\cE_3,\sa2\geq 0}{\min}\ \
&\left(\frac{1}{\sHd2}+\frac{1}{\sw2 }\frac{f_1(\cE_0,\alpha^2,\cE_2,\cE_3)}
{f_2(\cE_0,\alpha^2,\cE_2,\cE_3)} \right)^{-1} \\
\text{subject to}\ \ \ \ \ \ \ \ &\frac{\cE_3}{\Nt}\frac{1}{\sG2(\Nt-\NL)\sa2+\sw2}\leq \frac{1}{\gamma}-\frac{1}{\sG2}\\
&\cE_0+\cE_1+\cE_2+\cE_3+(\Nt-\NL)\sa2 \Nt\leq P_{ave}(\Nt+\Nt+\NL+\Nt) \\
&\cE_0+\cE_3+(\Nt-\NL)\sa2 \Nt\leq \bar{P}_t(\Nt+\Nt)\\
&\cE_1+\cE_2\leq \bar{P}_L(\Nt+\NL)
\end{align}
\end{subequations}
where
\begin{align*}
f_1(\cE_0,\alpha^2,\cE_2,\cE_3)=
\frac{\cE_3}{\Nt}\left(\NL \left(\frac{\sHd2 \cE_0}{\Nt}+\sw2\right)\alpha^2
+\frac{\Nt\sHu2}{\swt2}\left(\frac{\sHd2 \cE_0}{\Nt}+\sw2\right)\alpha^2\frac{\cE_2}{\NL}
+\frac{\cE_2}{\NL}
+\frac{\swt2}{\sHu2}\right)
\end{align*}
and
\begin{align*}
f_2(\cE_0,&\alpha^2,\cE_2,\cE_3)\\=&
(\Nt-\NL)\sa2\sHd2 \left(\frac{\NL}{\sw2} \left(\frac{\sHd2 \cE_0}{\Nt}+\sw2\right)\alpha^2
+\frac{\Nt\sHu2}{\swt2}\alpha^2\frac{\cE_2}{\NL}
+\frac{\cE_2}{\NL\sw2}
+\frac{\swt2}{\sHu2\sw2} \right)\notag\\
&+\left(\NL \left(\frac{\sHd2 \cE_0}{\Nt}+\sw2\right)\alpha^2
+\frac{\Nt\sHu2}{\swt2}\left(\frac{\sHd2 \cE_0}{\Nt}+\sw2\right)\alpha^2\frac{\cE_2}{\NL}
+\frac{\cE_2}{\NL}
+\frac{\swt2}{\sHu2}\right).
\end{align*}
By introducing the auxiliary variable
\begin{equation}\label{eq.auxi.t}
t=\frac{f_1(\cE_0,\alpha^2,\cE_2,\cE_3)}{f_2(\cE_0,\alpha^2,\cE_2,\cE_3)}
\end{equation}
and by defining the variables
$t_0=\frac{\sHd2 \cE_0}{\Nt}+\sw2$, $t_1=\alpha^2$, $t_2=\frac{\cE_2}{\NL}$, $t_3=\frac{\cE_3}{\Nt}$, and
$t_4=(\Nt-\NL)\sa2\sG2+\sv2$, the problem can be reformulated as
\begin{subequations}\label{prob.Nonrecip.3}
\begin{align}
\underset{t,t_0,t_1,t_2,t_3,t_4\geq 0}{\min}\ \
&t^{-1} \\
\text{subject to}\ \ \ \ \ \ \ \ &t\leq  \label{eq.auxi.t.ineq} \frac{\bar{f}_1(t_0,t_1,t_2,t_3)}{\bar{f}_2(t_0,t_1,t_2,t_3)}\\
&\sw2 t_0^{-1}\leq 1 \label{eq.constra.c}\\
&\sv2 t_4^{-1}\leq 1 \label{eq.constra.d}\\
&c_1t_3t_4^{-1}\leq 1 \label{eq.constra.e}\\
&c_2\left(\frac{\Nt}{\sHd2}t_0+\Nt\NL t_0t_1+\NL t_2
+\Nt t_3 +\frac{\Nt}{\sG2}t_4  \right)\leq 1 \\
&c_3\left(\frac{\Nt}{\sHd2}t_0+\Nt t_3 +\frac{\Nt}{\sG2}t_4  \right)\leq 1 \\
&c_4(\Nt\NL t_0t_1+\NL t_2)\leq 1 \label{eq.constra.h}
\end{align}
\end{subequations}
where
\begin{align*}
\bar{f}_1(t_0,t_1,t_2,t_3)=&
\NL t_0t_1t_3
+\frac{\Nt\sHu2}{\swt2}t_0t_1t_2t_3
+t_2t_3
+\frac{\swt2}{\sHu2}t_3,
\end{align*}
\begin{align*}
\bar{f}_2(t_0,t_1,t_2,t_3)=&
\left( \frac{t_4}{\sG2}-\frac{\sv2}{\sG2}\right)\sHd2\left(\frac{\NL}{\sw2}t_0t_1
+\frac{\Nt\sHu2}{\swt2}t_1t_2
+\frac{1}{\sw2}t_2
+\frac{\swt2}{\sHu2\sw2} \right)\notag \\
&+\NL t_0t_1
+\frac{\Nt\sHu2}{\swt2}t_0t_1t_2
+t_2
+\frac{\swt2}{\sHu2}
\end{align*}
and
\begin{align*}
c_1&=\left(\frac{1}{\gamma}-\frac{1}{\sG2}\right)^{-1},\ \
c_2=\left(P_{ave}(3\Nt+\NL)+\frac{\Nt\sw2}{\sHd2}+\frac{\sv2 \Nt}{\sG2} \right)\\
c_3&=\left(2\bar{P}_t\Nt+\frac{\Nt\sw2}{\sHd2}+\frac{\sv2 \Nt}{\sG2} \right),\ \
c_4=\left(\bar{P}_L(\Nt+\NL)\right)^{-1}.
\end{align*}
Note that in (\ref{eq.auxi.t.ineq}) the equality was replaced
by the inequality since one can inspect that the inequality
must be active when the optimal objective value is achieved.
To make sure $\cE_0$ and $\sa2$ are no less than zero, we attach
two posynomial constraints (\ref{eq.constra.c}) and (\ref{eq.constra.d}).
In addition, we have reformulated the constraints in
(\ref{eq.constra.e}-\ref{eq.constra.h}) into posynomial
inequalities, which are standard inequality constraints for GP.
However, the inequality constraint in (\ref{eq.auxi.t.ineq}) is
not a standard GP inequality. It can only be expressed as a
ratio of posynomials as given below:
\begin{equation}\label{eq.auxi.t.ineq.2}
\frac{\frac{\sHd2}{\sG2}\!\left(\!
\frac{\NL}{\sw2}t_0t_1t_4t
+\frac{\Nt\sHu2}{\swt2}t_1t_2t_4t
+\frac{1}{\sw2}t_2t_4t
+\frac{\swt2}{\sHu2\sw2}t_4t\!\right)\!
+\NL t_0t_1t
+\frac{\Nt\sHu2}{\swt2}t_0t_1t_2t
+t_2t
+\frac{\swt2}{\sHu2}t}
{\frac{\sv2\sHd2}{\sG2}\!\left(\!
\frac{\NL}{\sw2}t_0t_1
+\frac{\Nt\sHu2}{\swt2}t_1t_2
+\frac{1}{\sw2}t_2
+\frac{\swt2}{\sHu2\sw2}\!\right)\!
+\NL t_0t_1t_3
+\frac{\Nt\sHu2}{\swt2}t_0t_1t_2t_3
+t_2t_3
+\frac{\swt2}{\sHu2}t_3}\leq 1
\end{equation}
In order to simplify the problem into a standard GP form, we apply the monomial approximation \cite{Tutorial_GP}
to transform this into a posynomial constraint. In particular, if a set of feasible points
$\{\bar{t}, \bar{t}_1, \bar{t}_2, \bar{t}_3, \bar{t}_4\}$ of problem
(\ref{prob.Nonrecip.3}) is given, the inequality in (\ref{eq.auxi.t.ineq.2})
can be replaced by the posynomial constraint given below:
\begin{equation}
\frac{\frac{\sHd2}{\sG2}\!\left(\!
\frac{\NL}{\sw2}t_0t_1t_4t
+\frac{\Nt\sHu2}{\swt2}t_1t_2t_4t
+\frac{1}{\sw2}t_2t_4t
+\frac{\swt2}{\sHu2\sw2}t_4t\!\right)\!
+\NL t_0t_1t
+\frac{\Nt\sHu2}{\swt2}t_0t_1t_2t
+t_2t
+\frac{\swt2}{\sHu2}t}
{g(\bar{t}, \bar{t}_1, \bar{t}_2, \bar{t}_3)
\left(\frac{t_0}{\bar{t}_0}\right)^{\theta_0}
\left(\frac{t_1}{\bar{t}_1}\right)^{\theta_1}
\left(\frac{t_2}{\bar{t}_2}\right)^{\theta_2}
\left(\frac{t_3}{\bar{t}_3}\right)^{\theta_3}}\leq 1
\end{equation}
where
\begin{align*}
g(\bar{t}, \bar{t}_1, \bar{t}_2, \bar{t}_3)=&
\frac{\sv2\sHd2}{\sG2}\left(
\frac{\NL}{\sw2}t_0t_1
+\frac{\Nt\sHu2}{\swt2}t_1t_2
+\frac{1}{\sw2}t_2
+\frac{\swt2}{\sHu2\sw2}\right)\\
&+\NL t_0t_1t_3
+\frac{\Nt\sHu2}{\swt2}t_0t_1t_2t_3
+t_2t_3
+\frac{\swt2}{\sHu2}t_3,
\end{align*}
\begin{align*}
\theta_0&=\frac{\frac{\sv2\sHd2\NL}{\sG2\sw2}\bar{t}_0\bar{t}_1
+\NL\bar{t}_0\bar{t}_1\bar{t}_3+\frac{\Nt\sHu2}{\swt2}\bar{t}_0\bar{t}_1\bar{t}_2\bar{t}_3}
{g(\bar{t}, \bar{t}_1, \bar{t}_2, \bar{t}_3)},
\end{align*}
$$\theta_1=\frac{\frac{\sv2\sHd2\NL}{\sG2\sw2}\bar{t}_0\bar{t}_1
+\frac{\sv2\sHd2\sHu2\Nt}{\sG2\swt2}\bar{t}_1\bar{t}_2
+\NL\bar{t}_0\bar{t}_1\bar{t}_3+\frac{\Nt\sHu2}{\swt2}\bar{t}_0\bar{t}_1\bar{t}_2\bar{t}_3}
{g(\bar{t}, \bar{t}_1, \bar{t}_2, \bar{t}_3)},$$
$$\theta_2=\frac{
\frac{\sv2\sHd2\sHu2\Nt}{\sG2\swt2}\bar{t}_1\bar{t}_2
+\frac{\sv2\sHd2}{\sG2\sw2}\bar{t}_2
+\frac{\Nt\sHu2}{\swt2}\bar{t}_0\bar{t}_1\bar{t}_2\bar{t}_3+\bar{t}_2\bar{t}_3}
{g(\bar{t}, \bar{t}_1, \bar{t}_2, \bar{t}_3)},$$
and
$$\theta_3=\frac{
\NL \bar{t}_0\bar{t}_1\bar{t}_3
+\frac{\Nt\sHu2}{\swt2}\bar{t}_0\bar{t}_1\bar{t}_2\bar{t}_3+\bar{t}_2\bar{t}_3
+\frac{\swt2}{\sHu2}\bar{t}_3}
{g(\bar{t}, \bar{t}_1, \bar{t}_2, \bar{t}_3)}$$
Hence, for a given set of feasible points
$\{\bar{t}, \bar{t}_1, \bar{t}_2, \bar{t}_3, \bar{t}_4\}$, the problem
in (\ref{prob.Nonrecip.3}) can be approximate by the following problem
\begin{subequations}\label{prob.Nonrecip.4}
\begin{align}
\underset{e_0,e_1,e_2,e_3,\sa2 \geq 0}{\min}\ \ &t^{-1} \\
\text{subject to}\ \ \ \ \ \ \ \
&\frac{\frac{\sHd2}{\sG2}\left(
\frac{\NL}{\sw2}t_0t_1t_4t
+\frac{\Nt\sHu2}{\swt2}t_1t_2t_4t
+\frac{1}{\sw2}t_2t_4t
+\frac{\swt2}{\sHu2\sw2}t_4t\right)}
{g(\bar{t}, \bar{t}_1, \bar{t}_2, \bar{t}_3)
\left(\frac{t_0}{\bar{t}_0}\right)^{\theta_0}
\left(\frac{t_1}{\bar{t}_1}\right)^{\theta_1}
\left(\frac{t_2}{\bar{t}_2}\right)^{\theta_2}
\left(\frac{t_3}{\bar{t}_3}\right)^{\theta_3}}\notag\\
&\ \ \ \ \ \ \ \ \ \ \ +\frac{\NL t_0t_1t
+\frac{\Nt\sHu2}{\swt2}t_0t_1t_2t
+t_2t
+\frac{\swt2}{\sHu2}t}
{g(\bar{t}, \bar{t}_1, \bar{t}_2, \bar{t}_3)
\left(\frac{t_0}{\bar{t}_0}\right)^{\theta_0}
\left(\frac{t_1}{\bar{t}_1}\right)^{\theta_1}
\left(\frac{t_2}{\bar{t}_2}\right)^{\theta_2}
\left(\frac{t_3}{\bar{t}_3}\right)^{\theta_3}}\leq 1\\
&\sw2 t_0^{-1}\leq 1\\
&\sv2 t_4^{-1}\leq 1\\
&c_1t_3t_4^{-1}\leq 1 \\
&c_2\left(\frac{\Nt}{\sHd2}t_0+\Nt\NL t_0t_1+\NL t_2
+\Nt t_3 +\frac{\Nt}{\sG2}t_4  \right)\leq 1 \\
&c_3\left(\frac{\Nt}{\sHd2}t_0+\Nt t_3 +\frac{\Nt}{\sG2}t_4  \right)\leq 1 \\
&c_4(\Nt\NL t_0t_1+\NL t_2)\leq 1
\end{align}
\end{subequations}
The problem then becomes a standard GP problem and
can be efficiently solved by a few simple computer softwares such as
\texttt{CVX} \cite{CVX_DCP}. The condensation method then proposes to repeat this process iteratively by
replacing the set of feasible points $\{\bar{t}, \bar{t}_1, \bar{t}_2, \bar{t}_3, \bar{t}_4\}$ in each iteration
with the optimal solution of (\ref{prob.Nonrecip.4}) obtained in the previous iteration. This process continues until no further improvements can be obtained.


\end{document}